\documentclass[a4paper,10pt]{article}
\pdfoutput=1
\usepackage{jcappub}
\usepackage{mathtools}
\usepackage{ulem}
\usepackage{bm}
\usepackage{xcolor}
\usepackage{orcidlink}
\usepackage{xcolor}

\renewcommand{\thesection}{\arabic{section}}

\def\laq{~\raise 0.4ex\hbox{$<$}\kern -0.8em\lower 0.62ex\hbox{$\sim$}~}
\def\gaq{~\raise 0.4ex\hbox{$>$}\kern -0.7em\lower 0.62ex\hbox{$\sim$}~}

\def\beq{\begin{equation}}
\def\eeq{\end{equation}}
\def\bea{\begin{eqnarray}}
\def\eea{\end{eqnarray}}

\title{Viscosity in Isotropic Cosmological Backgrounds in General Relativity and Starobinsky Gravity}

\author[a]{Eliseo Pavone \orcidlink{0000-0002-3022-4545},}
\author[a] {Luigi Tedesco \orcidlink{0000-0001-6508-2658}}    

\affiliation[a]{Dipartimento di Fisica, Universit\`a di Bari, Via G. Amendola 173, 70126 Bari, Italy, and Istituto Nazionale di Fisica Nucleare, Sezione di Bari, Italy}
\emailAdd{eliseo.pavone@ba.infn.it}
\emailAdd{luigi.tedesco@ba.infn.it}

\abstract{We present a general analysis of the role of shear viscosity in cosmological backgrounds, focusing on isotropic space-time in both Einstein and $f(R)$ gravity. By computing the divergence of the stress-energy tensor in a general class of isotropic (but not necessarily homogeneous) geometries, we show that shear viscosity does not contribute to the background dynamics when the fluid is comoving. This result holds in both the Jordan and Einstein frames, and implies that shear viscosity cannot affect the electromagnetic luminosity distance which is determined by the background  light-like geodesics.
As an application of our results, we critically examine recent claims that shear viscosity can alter the Hubble evolution and the electromagnetic luminosity distance in Starobinsky gravity. We demonstrate that the continuity equation used in that work is at odds both with the covariant conservation of the stress-energy tensor and the local second law of thermodynamics. We further show that even in models where such modifications could mimic bulk viscosity, the resulting entropy evolution is inconsistent with standard thermodynamic expectations.}

\keywords{General Relativity, Viscous cosmology, Modified Gravity, Luminosity Distance, Gravitational Entropy}

\begin{document}
\maketitle
\vskip 0.8 cm

\section{Introduction}
In the last years the study of the gravitational and electromagnetic luminosity distance has been considered, because it is an excellent test for general relativity, in which the two distances are equal in Friedmann-Lamaître-Robertson-Walker (FLRW) backgrounds \cite{{Belgacem:2017ihm},{Maggiore:2018sht},{Belgacem:2018lbp},{LISACosmologyWorkingGroup:2019mwx},{Mukherjee:2020mha},{Begnoni:2024tvj},{Pantiri:2024bao},{Fonseca:2023uay},{Lizardo:2021ygl},{Garcia:2021svl},{Calcagni:2019kzo},{Bertacca:2017vod},{Yoo:2016vne},{Deffayet:2007kf},{Saltas:2014dha}}. In relation to this topic, 
in a recent paper \cite{Chi2025}, the authors expand on our previous results \cite{Fanizza_2022,Fanizza:2020hat,Fanizza:2021ngq} on the ratio between the gravitational wave luminosity distance and electromagnetic luminosity distance in a viscous dark matter cosmological model in Einstein and Starobinsky gravity \cite{STAROBINSKY198099,Nojiri:2010wj,Nojiri:2017ncd,Cognola:2005de}. They claim that  the presence of shear viscous fluids in Starobinsky gravity  is able to change the electromagnetic luminosity distance $d_L^\text{EM}$, and they perform cosmological parameter estimation using Supernova Ia, Cosmic Chronometers and Gamma Ray Bursts with two type of evolution for the shear viscosity $\eta$.\\
However we find that the very first assumption on the influence of the shear viscosity on the background evolution contradicts our previous results obtained in \cite{Fanizza_2022}, where we claimed that in a standard general relativistic $\Lambda$CDM cosmology,  shear viscosity does not modify the Friedmann equations and, leads to a luminosity distance for electromagnetic waves unaltered by its presence. This feature is preserved also in modified gravity theories, such as Starobinsky gravity.\\

The motivation for this paper is to point from general geometrical arguments that the time component of the divergence of the stress-energy tensor  $\nabla_\mu T^{\mu\nu}$, regardless of the gravitational theory under exam, cannot be affected by the presence of anisotropic stresses (that is the case for comoving viscous fluids), for a larger class of isotropic but non-homogeneous space-time that includes the FLRW case (Sect. 2), both in the Einstein Frame and in the Jordan Frame (Sect. 3) \cite{Bahamonde:2016wmz,Bahamonde:2017ize,Capozziello:2005mj,Odintsov:2022grf}, where we show how viscous coefficients and the continuity equation transform under conformal rescaling. \\
We perform  explicit evaluations that are at odds with the assumption of the dependence  of the Hubble parameter by shear viscous contributions of the dark matter fluid (Eq.(41)) of \cite{Chi2025}) that is to say
\begin{equation}
\dot{\rho}_{vdm}+3H\left(\rho_{vdm}+2\eta H\right)=0,\label{41}
\end{equation}
and as a consequence,  since   $d_L^\text{EM}$ is only related to  null geodesics of the background geometry via the well known relation \cite{Fanizza:2020hat}
\begin{equation}
d_L^\text{EM}=(1+z)\int_0^zdz'\frac{1}{H(z')}.\label{dl}
\end{equation}
Since, as we demonstrate below (see Eq.\eqref{continuitycorrect}), the Hubble parameter $H$ is independent of $\eta$ in both Starobinsky and Einstein gravity, the parameter estimation in \cite{Chi2025} rests on an assumption that does not hold in these frameworks; consequently, the resulting constraints on shear viscosity and the associated cosmological parameters may need to be re-evaluated. \\ Proving that shear viscosity does not alter the continuity equation, hence the evolution of the Hubble parameter $H$, we conclude that the parameter estimation of \cite{Fanizza:2020hat} still holds, even if shear viscosity is present. Constraints on shear viscosity in isotropic cosmologies cannot be derived from the luminosity distance alone and multi messenger events are required, as done in \cite{Fanizza_2022}, where the constraint on shear viscosity was obtained from combined observations of the standard siren GW170817 \cite{PhysRevLett.119.161101} and electromagnetic counterpart of its host galaxy NGC4993 \cite{Cantiello:2018ffy}. Under further inspection we find that the cosmological parameter estimation performed in \cite{Chi2025} is still reliable, but not to give estimation of shear viscosity, but on bulk, under the formal identification $2\eta\to-3\zeta$, where $\zeta$ represents the bulk viscosity. However the parameter estimation performed in \cite{Chi2025} finds a best fit with $\eta>0$ that implies a negative $\zeta$ posing questions on the consistency of the parameter best-fit in light of the second law of thermodynamics. \\
To this aim in Sect.4 we will show under what conditions a bulk viscous fluid violates the second law of thermodynamics both in comoving volumes and in an Hubble volume in light of the Generalized Second Law (GSL), where the apparent horizon entropy must be taken into account.  We conclude that the matter entropy in a Hubble volume is always decreasing in the $\Lambda$ dominated epoch, both in GR and Starobinsky model, even though this is a feature also for non viscous dust fluids in the standard $\Lambda$CDM model. We find that the total entropy, matter + horizon is always increasing, making GSL not a valuable tool to discriminate physically sensible  viscous modifications of dust fluid continuity equation both in GR and Starobinsky gravity.\\

\textit{Conventions}: \\
we use the metric signature $g_{\mu\nu}=\text{diag}(+,-,-,-)$,the Christoffel connection as $\Gamma_{\mu\nu}\,^{\alpha}=\frac{1}{2}g^{\alpha\rho}(\partial_\mu g_{\nu\rho}+\partial_\nu g_{\mu\rho}-\partial_\rho g_{\mu\nu})$ and the covariant derivative as $\nabla_\mu A_{\alpha\dots}=\partial_\mu A_{\alpha\dots}-\Gamma_{\mu\alpha}\,^{\nu} A_{\nu\dots}-\dots$, $c=\hbar=1$ and $\lambda_p^2=8\pi G$.

\section{The continuity equation in a class of isotropic geometries}
In this section we demonstrate that regardless of the gravitational theory under study, the covariant derivative of the stress-energy tensor for a comoving dissipative fluid, in a particular class of isotropic, but non-homogeneous space-time is independent from the shear viscosity. We work in the Eckart reference frame, where the 4-velocity is taken to be parallel to the particle flux and heat transfer is manifest. The  4-velocity can be expressed as $u^{\mu}(x)$ and it is a time-like vector $u_\mu u^\mu=1$. We can decompose the energy-momentum tensor in terms of projections parallel and perpendicular to $u^\mu$ by defining the following projector $h_{\mu\nu}=g_{\mu\nu}-u_\mu u_\nu$, with the  properties $h_{\mu\nu}h^{\mu\nu}=3$, $h_{\mu\alpha}h^{\alpha\nu}=\delta_\mu^\nu$ and orthogonal to the 4-velocity, $h_{\mu\nu}u^\nu=0$. In this  frame we identify the energy density $\rho$, the pressure $p$, the bulk pressure $\Pi$, the heat flux $q_\mu$ and  the anisotropic stress $\pi_{\mu\nu}$ (which is related to the shear viscosity), as follows \cite{Muronga:2003ta}
\begin{equation}
\begin{aligned}
    \rho&=T_{\mu\nu}u^\mu u^\nu, \\
    p+\Pi&=-\frac{1}{3} T_{\mu\nu} h^{\mu\nu},\\
    q_{\mu}&=h_\mu\,^\lambda u^\nu T_{\nu\lambda},\\
    \pi_{\mu\nu}&=T_{\langle \mu \nu\rangle},\label{geometricdec}
    \end{aligned} 
\end{equation}
where the angular parenthesis denote symmetric and tracless part of the tensor projected on the space-like sub-space orthogonal to $u_\mu$, namely
\begin{equation}
A_{\langle \mu \nu\rangle}=\big(h_\mu\,^{(\alpha|}h_\nu\,^{|\beta)}-\frac{1}{3}h_{\mu\nu}h^{\alpha\beta}\big)A_{\alpha\beta},\label{angular}
\end{equation}
where we denote the usual symmetrization in round brackets $B_{(\mu\nu)}\equiv\frac{1}{2}\left( B_{\mu\nu}+B_{\nu\mu}\right)$. With this typical conventions, the stress-energy tensor can be written as
\begin{equation}
    T_{\mu\nu}=\rho u_\mu u_\nu-(p+\Pi) h_{\mu\nu}+2u_{(\mu}q_{\nu)}+\pi_{\mu\nu}. \label{energy} 
\end{equation}
The geometrical meaning of each term is now manifest, every term in the stress-energy tensor is decomposed in the basis defined by the  projector tensor $h_{\mu\nu}$, the 4-velocity $u_\mu u_\nu$ and the vectorial part obtained by projecting the stress-energy tensor on $h_\mu\,^\lambda u^\alpha$. Most importantly  the anisotropic stress is the part of the stress-energy tensor, orthogonal to $u_\mu$ and traceless, so $u^\mu\pi_{\mu\nu}=0=h^{\mu\nu}\pi_{\mu\nu}$ and for this reason it is the source of the modification in the evolution equation of transverse and traceless tensor perturbations \cite{Fanizza:2021ngq}.  \\
 We also recall that the bulk pressure $\Pi$ and the anisotropic stress $\pi_{\mu\nu}$ are related via the bulk ($\zeta\geq0$) and shear ($\eta\geq0$) viscosity to the expansion rate $\Theta$  and shear tensor $\sigma_{\mu\nu}$ respectively, by the following relations
\begin{equation}
\begin{aligned}
    \Pi&\equiv-\zeta \Theta=-\zeta h^{\alpha}_{\mu}\nabla_\alpha u^\mu=-\zeta \nabla_\alpha u^\alpha,\\
    \pi_{\mu\nu}&\equiv2\eta \sigma_{\mu\nu}=2\eta h^\alpha_{\langle\mu|} \nabla_\alpha u_{|\nu\rangle}=2\eta \nabla_{\langle\mu|}u_{|\nu\rangle},\label{viscosity}
    \end{aligned}
\end{equation}
where in the last two equalities we used $u_\mu \nabla_\nu u^\mu=0$ and, the last term in the previous expression is the traceless and orthogonal to $u^\mu$ part of the strain-rate tensor $\nabla_{\mu}u_\nu$.
After some algebraic manipulation the stress-energy tensor can be expressed as
\begin{equation}
    \begin{aligned}
T_{\mu\nu}=&\rho u_\mu u_\nu-(p-\zeta \nabla_\alpha u^\alpha) h_{\mu\nu}+2u_{(\mu}q_{\nu)}\\
-&\eta\big(u_\mu u^\alpha \nabla_\alpha u_\nu+u_\nu u^\alpha \nabla_\alpha u_\mu-\nabla_\mu u_\nu-\nabla_\nu u_\mu +\frac{2}{3}h_{\mu\nu}\nabla_\alpha u^\alpha\big),\label{finalstress}
    \end{aligned}
\end{equation}
in concordance with \cite{Chi2025,Fanizza:2021ngq,Fanizza_2022}.
Now we are going to compute the divergence of  Eq.~\eqref{finalstress}. In order to have the energy continuity  equation,  we evaluate the projection of the divergence of the stress-energy tensor on the 4-velocity, $u_\mu\nabla_\nu T^{\mu\nu}$. Using the following identities
\begin{equation}
u_\mu q^\mu=0, \qquad \nabla_\mu h^{\mu\nu}=-u^\nu \Theta, \qquad u_\mu \nabla_\nu q^\mu=-q^\mu \nabla_\nu u_\mu,
\end{equation}
and the orthogonality relations, we have 
\begin{equation}
    \begin{aligned}
        u_\mu \nabla_\nu(\rho u^\mu u^\nu)&=u^\nu\nabla_\nu\rho +\rho \Theta,\\ 
        u_\mu \nabla_\nu[ (p+\Pi)h^{\mu\nu}]&=-(p+\Pi)\Theta,\\
        u_\mu\nabla_\nu(2q^{(\mu}u^{\nu)})&=\nabla_\nu q^\nu-u^\nu q^\mu \nabla_\nu u_\mu,\\
        u_\mu\nabla_\nu\pi^{\mu\nu}&=\eta\big[-(\nabla_\nu u^\alpha)(\nabla_\alpha u^\nu)-(\nabla^\alpha u_\nu)(\nabla_\alpha u^\nu)+ u^\nu u^\alpha(\nabla_\nu u_\mu)(\nabla_\alpha u^\mu)+\frac{2}{3}\Theta^2\big]\\
        &=-2\eta\sigma^2,   
    \end{aligned}\label{divergence}
\end{equation}
where in the last equality we have defined $\sigma^2\equiv\sigma_{\mu\nu}\sigma^{\mu\nu}$. Finally using Eqs.~\eqref{divergence} we find 
\begin{equation}
u_\mu\nabla_\nu T^{\mu\nu}=u^\nu\nabla_\nu\rho +(\rho+p+\Pi) \Theta+\nabla_\nu q^\nu-u^\nu q^\mu \nabla_\nu u_\mu-2\eta \sigma^2.\label{final}
\end{equation}
The question whether or not the shear viscosity can enter the energy equation is now a matter of evaluating the shear tensor for a given geodesic in a given space-time.\\
For the sake of generality we do not assume homogeneity for our background geometry, namely we use a rotationally invariant (isotropic) metric in the synchronous gauge, with line element given by:
\begin{equation}
ds^2=dt^2-e^{2\lambda(t,r)}\left(A(r)^2dr^2+B(r)^2 d\Omega_2^2\right),\label{lineel}
\end{equation}
with $d\Omega_2^2=d\theta^2+\sin^2(\theta)d\phi^2$. The reason behind this choice is to prove, as long as the fluid is comoving, that shear viscosity cannot enter into the expression $u_\mu\nabla_\nu T^{\mu\nu}$  in more generic space-time that are isotropic but not homogeneous.  We would also like to stress, that for the metric given in Eq.~\eqref{lineel}, a comoving observer, $u^\mu=(1,0,0,0)$ is also geodesic, $u^\mu \nabla_\mu u^\nu=\Gamma_{00}\,^{\nu}=0$, so we will assume that our fluid is comoving  with the geometry itself.\\
Under the assumption of comoving fluid it is straight forward to evaluate the shear tensor
\begin{equation}
\begin{aligned}
\sigma_{\mu\nu}&=\nabla_{(\mu}u_{\nu)}-u_{(\mu|} u^\alpha \nabla_\alpha u_{|\nu)}-\frac{1}{3}h_{\mu\nu}\nabla_\lambda u^\lambda\\
&=\nabla_{(\mu}u_{\nu)}-\frac{1}{3}h_{\mu\nu}\nabla_\lambda u^\lambda,
\end{aligned}\label{shear}
\end{equation}
where in the last line we used the geodesic equations. The $0-\mu$ component of Eq.~\eqref{shear} is vanishing because of the orthogonality conditions $\sigma_{\mu0}= u^\nu \sigma_{\mu\nu}=0$, while the $i-j$ components are given by
\begin{equation}
    \sigma_{ij}=-\Gamma_{ij}\,^0-\frac{1}{3}g_{ij}\Gamma_{\lambda0}\,^\lambda.\label{final2}
\end{equation}
Defining $\partial_0 \lambda(t,r)\equiv H(t,r)$, we have $\Gamma_{ij}\,^0=-H(t,r)g_{ij}$ and $\Gamma_{\lambda0}\,^\lambda=3H(t,r)$, so $\sigma_{ij}=0$ and  we conclude that the shear tensor is vanishing and so $\sigma^2=0$. This is one of the main results of this paper and stands in contrast with the expression proposed Eq.~\eqref{41}. We generically proved that the gravitational equations of a yet to be specified gravitational theory, cannot be affected by the presence of a shear viscous contribution, as long as the background fluid source is comoving.\footnote{Assuming that the gravitational equations are in the form $L_{\mu\nu}=\lambda_p^2 T_{\mu\nu}$, where the tensor $L_{\mu\nu}$ is intended to be containing the geometrical information of the theory (e.g. $L_{\mu\nu}=G_{\mu\nu}$ in General Relativity)}.  As a consequence the background evolution is not affected by its presence and, so are null geodesics, implying that also the luminosity distance is independent on the shear viscosity. It is easy to check that upon the following identifications, $e^{2\lambda(t,r)}\to a^2(t)$, $A(r)^2\to \frac{1}{1-kr^2}$ and $B(r)^2\to r^2$, the line element Eq.~\eqref{lineel} is the standard FLRW one. The immediate consequence is that no shear contribution can be present in the energy equation, nor in the background evolution equations, hence it cannot affect the background geometry evolution.\\
For completeness we evaluate Eq.~\eqref{final} for a comoving fluid in the space-time given by Eq.~\eqref{lineel}, including possible heat flux, $q^\mu\neq0$.  Due to the choice of the metric and the comoving fluid, the only component of the heat flux that is non trivial is $q^1$. This statement can be understood using the orthogonality $q^\mu u_\mu=0$, which implies $q^0=0$ and, the Einstein (Starobinsky) equations. Defining
\begin{equation}
\mathcal{A}_{\mu\nu}\equiv2R\,R_{\mu\nu}+g_{\mu\nu}\big(2\nabla^2 R-\frac{1}{2}R^2\big)-2\nabla_\mu\nabla_\nu R,\label{staro}
\end{equation}
we can write the Einstein (Starobinsky) equations as
\begin{equation}
G_{\mu\nu}+\alpha \mathcal{A}_{\mu\nu}=\lambda_p^2 T_{\mu\nu},\label{einstaro}
\end{equation}
where $G_{\mu\nu}$ is the Einstein tensor $G_{\mu\nu}\equiv R_{\mu\nu}-\frac{1}{2}g_{\mu\nu}R$ and $\alpha$ is a constant. The only off diagonal terms in these equations are the one given by the $0-1$ component and they can be sourced only by the term $q_1 u_0$ in the stress-energy tensor. The other off-diagonal terms are vanishing on the left hand side of Eq.~\eqref{einstaro} and enforce $q_\mu=(0,q_1,0,0)$\footnote{The presence in General Relativity ($\alpha=0$) of a trivial heat flux $q^{\mu}=0$, would imply $G_{01}=-2\partial_r H(t,r)=0$, and would enforce the separability $e^{\lambda(t,r)}=a(t)C(r)$. The factor $C(r)$ can be absorbed by a redefinition of the functions $A(r)$ and $B(r)$ leading to a more restricted space-time than the one in  Eq.~\eqref{lineel}.}. Denoting with a dot the derivative with respect to $t$ and with a $'$ the derivative with respect to $r$, we find that Eq.~\eqref{final} reduces to 
\begin{equation}
u_\mu\nabla_\nu T^{\mu\nu}=\dot{\rho}+q'^{1} +3H(t,r)\left[\rho+p-3\zeta H(t,r)\right]+3\lambda'(t,r)q^1+\big(\frac{A'(r)}{A(r)}+2\frac{B'(r)}{B(r)}\big)q^1. \label{generalcorrect}
\end{equation}
Assuming, $\zeta=0$, $q^{\mu}=0$, a dust dark matter $p=0$, and an FLRW geometry, namely $H(t,r)=H(t)$, recalling that as consequence of the Bianchi identities, both in Einstein and Starobinsky theories the divergence of the left hand side of Eq.~\eqref{einstaro} is vanishing \cite{Fanizza:2020hat}, 
 we recover from Eq.~\eqref{generalcorrect} the standard continuity equation (we denote with $\rho_{vdm}$ the energy density of viscous dark matter)
\begin{equation}
u_\mu\nabla_\nu T^{\mu\nu}=\dot{\rho}_{vdm}+3H(t)\rho_{vdm}=0. \label{continuitycorrect}
\end{equation}
We emphasize that no shear viscous contributions are present, which stands in contrast with the expression proposed in Eq.(41) of \cite{Chi2025}. However, if we allow the presence of a non trivial bulk viscosity $\zeta\neq0$, but retaining the FLRW geometry we find from Eq.~\eqref{generalcorrect} the well known result
\begin{equation}
   u_\mu\nabla_\nu T^{\mu\nu}=\dot{\rho}_{vdm}+3H(t)\big[\rho_{vdm}-3\zeta H(t)\big]=0.  \label{bulk}
\end{equation}
This entails that the effect captured in the parameter estimation of \cite{Chi2025} has to be regarded as arising from a bulk viscous contribution, indeed one can check that Eq.~\eqref{41} is formally identical to Eq.~\eqref{bulk}, under the following formal identification $2\eta\to-3\zeta$, which suggests that the effective parameter fitted in the above mentioned paper may actually correspond to a bulk viscous contribution. However the best-fit performed leads to a positive shear viscosity, which under this identification would imply a negative bulk viscosity, in tension with the usual thermodynamic requirement $\zeta \geq 0$. An in-depth thermodynamic analysis of this point will be carried out in Sect. 4.\\ 
We note that the analysis has been carried out without assuming any specific gravitational theory, and as a general feature, even for the broader class of isotropic geometries considered in Eq.~\eqref{lineel}, the covariant divergence of the stress-energy tensor projected along the comoving 4-velocity does not include contributions from anisotropic stress. \\
In $f(R)$ gravity, this conclusion remains valid even when reformulating the theory in the so-called Einstein frame: no shear contributions appear in the continuity equation. However, additional terms can arise on the right-hand side of Eq.~\eqref{continuitycorrect} due to the conformal rescaling of the metric $g_{\mu\nu}^J = e^{\psi} g_{\mu\nu}^E$. These include terms like $2\rho\dot{\psi}$ in the Jordan frame or $\frac{\dot{\psi}}{2}(\rho - 3p)$ in the Einstein frame, depending on the comoving frame adopted. While these terms may resemble the structure of viscous contributions, their form is not arbitrary: they must be consistent with the Friedmann equations and the dynamics of a minimally coupled scalar field with potential $\tilde{V}(\psi) = \frac{Rf'(R) - f(R)}{f'(R)^2}$ and $\psi = -\ln f'$. In the next section, we will derive the continuity equation for a dissipative fluid in both frames for the general isotropic space-time given in Eq.~\eqref{lineel}.

\section{Conformal rescaling and the continuity equation: absence of shear viscous contributions in general isotropic space-time}

In this section, we provide a detailed derivation to support our claim that the modification of the continuity equation under conformal rescaling is not associated with shear viscosity effects. Although  modifications of the continuity equation arise naturally in the Einstein frame (EF) when matter is non-minimally coupled via the conformal factor, they cannot be interpreted as contributions from shear viscosity. This holds regardless of the specific gravitational theory considered, as long as the Einstein frame exists. One can foresee without any explicit computation that based solely on geometrical arguments, a conformal rescaling can alter ``volume changing" terms, but not  deformation terms such as the shear tensor for the very nature of conformal transformations, since they are volume changing and angle preserving.  It is worth noting that in the Jordan frame, where conformal rescaling is not active, such modifications do not occur. Even going from the Jordan frame (JF) to the Einstein frame (EF), shear viscosity only couples to the shear tensor,  that is vanishing in both frames when an FLRW space-time is assumed, or the more general class considered in Eq.~\eqref{lineel}. In the following we are going to give a detailed analysis of this statement. \\
Starting with a minimally coupled dissipative fluid in the JF we perform a conformal transformation in the form $g_{\mu\nu}^J\to e^{\psi} g_{\mu\nu}^E$, without specifying the gravitational theory under exam. For instance the conformal factor $\psi$ can be identified as the dilaton field $\phi$ when going from the so called String frame to the Einstein frame while dealing with the effective tree-level action of the bosonic massless multiplet of string theory in $4$-dimensions
\begin{equation}
S^S_\text{string}\sim -\int d^4x \sqrt{-g^S}e^{-\phi}(R^S + (\nabla\phi)^2-\frac{1}{12}H_{\mu\nu\alpha}H^{\mu\nu\alpha})+S^S_m(g^S_{\mu\nu},...),
\end{equation}
that after conformal rescaling $g_{\mu\nu}^S\to e^{\phi} g_{\mu\nu}^E$ becomes 
\begin{equation}
S^E_\text{string}\sim -\int d^4x \sqrt{-g^E}(R^E -\frac{1}{2} (\nabla\phi)^2-\frac{1}{12}e^{-2\phi}H_{\mu\nu\alpha}H^{\mu\nu\alpha})+S^E_m(g^E_{\mu\nu}e^{\phi},...).
\end{equation}
In $f(R)$ gravity the action can be recasted in a scalar-tensor form via the introduction of a Lagrange mulptiplier $\chi=f'(R)$
\begin{equation}
S^J=-\frac{1}{2\lambda_P^2} \int d^4x \sqrt{-g^J}(\chi R -V(\chi))+ S_m(g_{\mu\nu}^J,...),
\end{equation}
where $V(\chi)=R f'(R)-f(R)$. A conformal rescaling of the metric and neglecting total derivatives  leads to
\begin{equation}
S^E=-\frac{1}{2\lambda_P^2}\int d^4x \sqrt{-g^E}e^{2\psi}\Big[e^{-\psi}\chi(R^E-\frac{3}{2}(\nabla\psi)^2)-V(\chi)\Big]+S_m(g_{\mu\nu}^Ee^{\psi},...).
\end{equation}
Fixing $\psi=-\ln\chi$  we end up with the following action
\begin{equation}
S^E= -\frac{1}{2\lambda_P^2}\int d^4x \sqrt{-g^E}(R^E-\frac{3}{2}(\nabla\psi)^2-\tilde{V}(\psi)\Big]+S_m(g_{\mu\nu}^Ee^{\psi},...),
\end{equation}
where $\tilde{V}(\psi)=\frac{Rf'(R)-f(R)}{f'(R)^2}$.
We emphasize that an exponential pre-factor arise in the matter lagrangian which couples non trivially the new scalar degree of freedom $\psi$ to matter, namely
\begin{equation}
S_m^E\sim\int d^4x \sqrt{-g^E} e^{2\psi}\mathcal{L}^J(g_{\mu\nu}^E e^\psi,...).
\end{equation}
This is the starting point of our analysis. We start with a minimally coupled fluid in the JF and under conformal rescaling we will derive the continuity equation in the EF, without specifying the particular theory under exam, (as long as a conformal rescaling can bring the original theory action to an Einsteinian form). We will show that also in this case, even if additional terms arise in the case of Starobinsky gravity, they do not couple to the shear tensor in the isotropic metric considered in Eq.~\eqref{lineel}, hence in the FLRW case.\\
The general transformation for the energy-momentum tensor can be evaluated by varying the matter action in the JF with respect to its metric and applying the conformal transformation (we evaluate the transformation in general $d+1$ dimensions)
\begin{align}
\delta_{g^J}S_{m}&\equiv \frac{1}{2}\int\,d^{d+1}x\sqrt{|g^J|}T^J_{\mu\nu}\delta g^{\mu\nu}_J\,=\frac{1}{2}\int\,d^{d+1}x\sqrt{|g^{E}|}e^{\frac{d-1}{2}\psi}T^J_{\mu\nu}\delta g^{\mu\nu}_{E}\,,\nonumber\\
\delta_{g^{E}}S_{m}&\equiv\frac{1}{2}\int\,d^{d+1}x\sqrt{|g^E|}T^{E}_{\mu\nu}\delta g^{\mu\nu}_{E},
\end{align}
so by comparing the last two lines we get
\begin{equation}
T^{J}_{\mu\nu}=e^{-\frac{d-1}{2}\psi}\,T_{\mu\nu}^E.\label{comparison}
\end{equation}
In the following we are going to assume that, as in $f(R)$ theories, the stress-energy tensor is conserved in the JF \footnote{this is not the case for the effective theory coming from string theory when there is a dilatonic charge $\delta_\phi S_m=-\frac{1}{2}\int d^{d+1}x \sqrt{|g|}\sigma\delta \phi\neq0$.}, and we are going to evaluate the divergence on both sides. We recall that the Christoffel connection transforms as follows
\begin{equation}
\Gamma^J_{\mu\nu}\,^{\alpha}=\Gamma^E_{\mu\nu}\,^{\alpha}+\mathcal{C}_{\mu\nu}\,^\alpha,\qquad\qquad
\mathcal{C}_{\mu\nu}\,^\alpha\equiv\frac{1}{2}\big(\delta_\nu^\alpha\partial_\mu\psi+\delta_\mu^\alpha\partial_\nu\psi-g_{\mu\nu}^Eg^{\alpha\rho}_E\partial_\rho\psi\big).\label{connection}
\end{equation}
Using the continuity equation in the JF we obtain the following identity in the EF
\begin{equation}
0=\nabla_\mu T^{\mu\nu}_J=\nabla^E_\mu\big(e^{-\frac{d+3}{2}\psi}T^{\mu\nu}_E\big)+\mathcal{C}_{\mu\alpha}\,^\mu e^{-\frac{d+3}{2}\psi}T^{\alpha\nu}_E+\mathcal{C}_{\mu\alpha}\,^\nu e^{-\frac{d+3}{2}\psi}T^{\mu\alpha}_E, \label{continuityEprel}
\end{equation}
where we defined $\nabla^E_\mu A^{\alpha\nu\dots}\equiv\partial_\mu A^{\alpha\nu\dots}+\Gamma^E_{\mu\rho}\,^{\mu}A^{\rho\nu\dots}+\dots$. A straight forward evaluation leads using Eqs.~\eqref{connection} to the conservation equation in the EF as 
\begin{equation}
   \nabla^E_\mu\,T^{\mu\nu}_E=\frac{T_E}{2}\partial^\nu\psi,\label{continuityfinE}
\end{equation}
where $T_E=g_{\mu\nu}^ET^{\mu\nu}_E$ is the trace of the stress-energy tensor in the Einstein frame. Following the geometrical definitions given in Eqs.~\eqref{geometricdec} and demanding that $u_\mu^Ju^\mu_J=1=u_\mu^Eu^\mu_E$, so that $u_J^\mu=e^{-\frac{\psi}{2}}u^\mu_E$, $h_{\mu\nu}^J=g_{\mu\nu}^J-u_\mu^Ju_\nu^J=e^\psi h_{\mu\nu}^E$ we find
\begin{equation}
\Pi^J=e^{-\frac{\psi}{2}}\Pi^E,\qquad \Pi^E\equiv-\zeta^J\big(\Theta^E+\frac{d}{2}u^{\mu}_E\partial_\mu\psi\big), \qquad \Theta^E\equiv\nabla^E_\mu u^\mu_E, \label{rateE}
\end{equation}
and recalling the definition of the angular parenthesis given in Eq.~\eqref{angular}  (in $d$ dimensions)
\begin{equation}
A_{\langle\mu\nu\rangle}=\big(h_\mu\,^{(\alpha|}h_\nu\,^{|\beta)}-\frac{1}{d}h_{\mu\nu}h^{\alpha\beta}\big)A_{\alpha\beta},
\end{equation}
we find for the anisotropic stress
\begin{equation}
\begin{aligned}
    \pi_{\mu\nu}^J&=2\eta^J\sigma^J_{\mu\nu}=2\eta^J e^{\frac{\psi}{2}}\sigma_{\mu\nu}^E\\
    \sigma_{\mu\nu}^E&\equiv\nabla^E_{\langle\mu}u^E_{\nu\rangle},\label{anisotropicE}
    \end{aligned}
\end{equation}
and finally $q_\mu^J=q^E_\mu e^{-\frac{d}{2}\psi}$. Using Eq.~\eqref{energy} we are now able to evaluate the stress-energy tensor in the E-frame using Eqs.~\eqref{rateE} and \eqref{anisotropicE} and inverting the relation \eqref{comparison}
\begin{equation}
T_{\mu\nu}^E=\rho^E u^E_\mu u^E_\nu-\big[p^E-\zeta^E\big(\Theta^E+\frac{d}{2} u^\mu_E \partial_\mu\psi\big)\big] h^E_{\mu\nu}+2u^E_{(\mu}q^E_{\nu)}+2\eta^E\sigma^E_{\mu\nu}\label{energyE}
\end{equation}
where we defined the following transformations laws from the JF to the EF
\begin{equation}
        \rho^E\equiv e^{\frac{d+1}{2}\psi}\rho^J,\qquad p^E\equiv e^{\frac{d+1}{2}\psi}p^J,\qquad\zeta^E\equiv e^{\frac{d}{2}\psi}\zeta^J,\qquad \eta^E\equiv e^{\frac{d}{2}\psi}\eta^J.\label{redefinitions}
\end{equation}
The divergence in the EF of the stress-energy tensor follows the same relations of Eqs.~\eqref{divergence} 
\begin{equation}
    \begin{aligned}
        u_\mu^E \nabla^E_\nu(\rho^E u^\mu_E u^\nu_E)&=u^\nu_E\nabla^E_\nu\rho^E +\rho^E \Theta^E,\\ 
        u^E_\mu \nabla^E_\nu[ (p^E+\Pi^E)h^{\mu\nu}_E]&=-(p^E+\Pi^E)\Theta^E,\\
        u^E_\mu\nabla^E_\nu(2q_E^{(\mu}u_E^{\nu)})&=\nabla^E_\nu q^\nu_E-u^\nu_E q^\mu_E \nabla^E_\nu u^E_\mu,\\
        u^E_\mu\nabla^E_\nu\pi^{\mu\nu}_E&=-2\eta^E\sigma^2_E,   
    \end{aligned}\label{divergenceE}
\end{equation}
and the trace of Eq.~\eqref{energyE} reads 
\begin{equation}
    T^E=\rho^E-d(p^E+\Pi^E)=\rho^E-dp^E+d\zeta^E\Theta^E+\frac{d^2}{2}u^\mu_E\partial_\mu\psi.
\end{equation}
Projecting on the four velocity the right hand side of Eq.~\eqref{continuityfinE} we finally obtain the continuity equation in the EF
\begin{equation}
\begin{aligned}
    &u^\nu_E\nabla^E_\nu\rho^E +(\rho^E+p^E+\Pi^E) \Theta^E-\frac{u^\mu_E\partial_\mu\psi}{2}\big[\rho^E-d(p^E+\Pi^E)\big]\\
    &+\nabla^E_\nu q^\nu_E
-u^\nu_E q^\mu_E \nabla_\nu^E u_\mu^E-2\eta^E \sigma^2_E=0.\label{finalcontE}
\end{aligned}
\end{equation}
This equation justifies the expectations we exposed at the  beginning, the conformal rescaling cannot affect the shear part of the stress-energy tensor if the stress tensor is vanishing in one frame, due to Eq.~\eqref{anisotropicE}.
To fully conclude our argument we have to specify a background metric and the motion of our background fluid.   Two possibilities now arise, the fluid is geodesic in the EF, or the fluid is geodesic in the JF. It is widely known that a geodesic fluid in the JF, under conformal rescaling is not geodesic in the EF in general and vice versa, and only a radiation fluid stays geodesics (i.e. conformal transformations preserve the light-like geodesics), so we are going to analyze both cases. As will become evident, modifications of the continuity equation consistent with shear viscosity do not naturally arise from the conformal structure, regardless of the frame considered.\\
\subsection{Geodesic fluid in the Jordan Frame}
Although one can go through an explicit derivation of the shear tensor in the EF, this is not necessary, since  we have already proven that for the general class of space-time Eq.~\eqref{lineel}, $0=\pi_{\mu\nu}^J=2\eta^J\sigma_{\mu\nu}^J$, but making use of the transformations in Eqs.~\eqref{anisotropicE}, also $\sigma_{\mu\nu}^E=0$, hence $\sigma_E^2=0$ and no shear viscous contributions are present in the continuity equation \eqref{finalcontE} also in the EF. Being valid for the general case of Eq.~\eqref{lineel}, it still holds for the particular case of FLRW. This result says that even from a  phenomenological stand point, the correction to the conservation equation due to the conformal rescaling, does not involve shear viscosity.  We recall that physical observables such as the luminosity distance are frame independent \cite{Deruelle:2010ht,Chiba:2013mha,Rondeau:2017qbi}, so no matter the frame, the observed luminosity distance is not affected by which frame one decide to work with.
 From the geodesic condition in the JF  $u^\nu_J\nabla_\nu^J(u^\mu_J)=0$ we have that $u^\nu_E\nabla_\nu^E(u^\mu_E)=\frac{1}{2}(\partial^\mu\psi-u^\mu_Eu^\nu_E\partial_\nu\psi)$ and from the comoving condition in the Jordan frame (we are fixing $d=3$ from now on), we have $u^\mu_E=e^{\frac{\psi}{2}}(1,0,0,0)$. From $q_\mu^Eu^\mu_E=0$ we have $q_0^E=0$,  and moreover  we know that the Einstein equations can be written as $G_{\mu\nu}=\lambda_P^2T^{tot}_{\mu\nu}$, where  $T^{tot}_{\mu\nu}=T_{\mu\nu}^\psi+T_{\mu\nu}^{fluid}$, where the stress-energy tensor for $\psi$ corresponds to the one of a scalar field with potential $\tilde{V}(\psi)$. From the symmetries of the metric under study, we know that also in the EF, where the metric is $g_{\mu\nu}^E=e^{-\psi}g_{\mu\nu}^J$, the only off-diagonal non-vanishing components of the Einstein tensor are the $0-1$, hence $q^\mu_E=(0,q^1_E,0,0)$ and $\psi=\psi(t,r)$. Using for the JF metric the one given in Eq.~\eqref{lineel}, we get the continuity equation as
\begin{equation}
\begin{aligned}
  &\dot{\rho}^E+3H(t,r)(\rho^E+p^E-3\zeta^EH(t,r)e^{-\psi})\\
  &-2\rho^E\dot{\psi}+e^{-\frac{\psi}{2}}\left[q'^1_E+q^1_E\left(\frac{A'}{A}+2\frac{B'}{B}+3\lambda'-\frac{3}{2}\psi'\right)\right]=0.
  \end{aligned}
\end{equation}
In the FLRW limit for dust fluid, where $\psi=\psi(t)$ and with vanishing bulk viscosity we get
\begin{equation}
\dot{\rho}^E+3H(\rho^E+p^E)-2\rho^E\dot{\psi}=0,\label{conformalfirstcase}
\end{equation}
which may look similar to Eq.~\eqref{41}, but has a profound different meaning. Indeed that coupling between the conformal field and the energy density arises due to the interactions between the fluid and the conformal field, when changing frame. This correction may look similar to Eq.~\eqref{41}, if one assumes (in the dust case $p=0$) $\rho^E\sim H^2$, and with $\dot{\psi}$ taking the role of their shear viscosity (with a reversed sign), but  the dynamical field $\psi$ is not arbitrary and must satisfy its equation of motion and be consistent with the Friedmann equations. This confirms our initial statement: in the Einstein frame, corrections induced by conformal transformations are constrained and cannot reintroduce shear contributions.
\subsection{Geodesic fluid in the Einstein Frame}
For this case we assume that the line element in the Einstein frame is the one given by Eq.~\eqref{lineel}  and that the fluid is comoving $u^{\mu}_E=(1,0,0,0)$, hence geodesic in the EF. For the same reasons given in the previous subsection, also in this case, the shear tensor vanishes identically, ruling out any shear-viscous contribution to the conservation equation. In this case the continuity equation reads
\begin{equation}
\begin{aligned}
  &\dot{\rho}^E+3H(t,r)\Big[\rho^E+p^E-3H(t,r)e^{-\frac{3}{2}\psi}\zeta^E\Big]\\
  &-\frac{\dot{\psi}}{2}\left(\rho^E-3p^E-\frac{3}{2}e^{-\frac{3}{2}\psi}\zeta^E\dot{\psi}\right)+q^1_E\left(\frac{A'}{A}+2\frac{B'}{B}+3\lambda'\right)=0,
  \end{aligned}
\end{equation}
and finally in the FLRW limit and for vanishing bulk viscosity we recover
\begin{equation}
    \dot{\rho}^E+3H\big(\rho^E+p^E\big)-\frac{\dot{\psi}}{2}\left(\rho^E-3p^E\right)=0. \label{conformalsecondcase}
\end{equation}
Also in this case we can conclude that a new term akin to the one of Eq.~\eqref{41} arises, if one assumes (recalling $p=0$ for dust) $\dot{\psi}\rho^E\sim-4\eta H^2$, but also in this case the evolution of the additional term is constrained by the equation of motion of the scalar field and from the Friedmann equations. The same conclusions as in the case of a fluid comoving in the JF can be drawn also in this case.\\
\section{Second law of thermodynamics in comoving volumes and Generalized Second Law (GSL)}
In this section we give some thermodynamical considerations on possible viscous corrections in the continuity equation and whether they can violate the second principle of thermodynamics. We find that the parameter estimation  in \cite{Chi2025} leads to a violation of the second law of thermodynamics  in any comoving volume, while Generalized Second Law (GSL) arguments do not provide any valuable constraint on viscous modifications of the continuity equation, both in GR and Starobinsky $\Lambda$CDM model.\\
\subsection{Violation of the Second law in comoving volumes}
We start by analyzing the entropy variation  in a comoving volume of fixed coordinate distance $r^*$. In fact it is easy to see that upon time differentiation of  the first law of thermodynamics in an FLRW geometry with scale factor $a(t)$, for a dust fluid ($p=0$), $TdS=dE$, we recover the usual relation 
\begin{equation}
T \frac{\dot{S}}{V}=\dot{\rho}_{vdm}+3H\rho_{vdm}, \label{thermo}
\end{equation}
where $V=(r^* a)^3$ and $H\equiv\frac{\dot{a}}{a}$ is the Hubble rate. It is immediate to see that in this case the non-dissipative dust continuity equation Eq.~\eqref{continuitycorrect} leads to $\dot S=0$, hence the comoving entropy is constant.
Inserting Eq.~\eqref{thermo} into Eq.~\eqref{41} we obtain
\begin{equation}
T\frac{\dot{S}}{V}=-6\eta H^2\leq0,\label{violation}
\end{equation}
and since $\eta\geq0$, the continuity equation Eq.~\eqref{41} clearly violates the second law of thermodynamics. Moreover even under the more conservative interpretation that the parameter estimation performed in \cite{Chi2025} has to be intended as effectively coming from a bulk viscous contribution $-2\eta\to3\zeta$ we find the known result 
\begin{equation}
    T\frac{\dot{S}}{V}=9\zeta H^2\geq0,\label{bulkentropy}
\end{equation}
which leads to an entropy increase. However for the mean given by the best-fit  found in \cite{Chi2025} we have $\eta>0$, which immediately translates to $\zeta<0$, hence a violation of Eq.~\eqref{bulkentropy} and a decrease in entropy.\\
We stress that, because the dark matter fluid is (by assumption) comoving with our chosen volume, one would expect dissipation to \textit{increase} entropy locally. Indeed, this is the usual expectation in viscous fluid dynamics, where an isolated viscous fluid's entropy production is positive. Yet, our calculation finds the opposite (a net decrease), signaling a breakdown of the normal expectation for a purely viscous fluid.\\
 Our thermodynamic analysis is based on comoving volumes in an FLRW background, where spatial homogeneity and isotropy are assumed. In this setting, the entropy in each comoving patch must not decrease unless dissipative processes are active. A decrease in matter entropy in one region would require a compensating increase elsewhere, implying an inhomogeneous entropy distribution, which contradicts the FLRW homogeneity assumption.
\\
\subsection{Generalized Second Law in viscous $\Lambda$CDM, GR case}
We are now going to analyze entropy variation within a sphere of Hubble radius,  which is a different possibility of choosing a reference volume. All the considerations below assume  that the fluid is in thermodynamic equilibrium with the horizon, $T_m=T_H$, that is to say that the matter content has the same temperature as the one of the horizon (see \cite{Paul:2025rqe, Odintsov:2024ipb, Nojiri:2025gkq}  for recent discussions on the emergence of bulk viscosity in cosmological fluids also when $T_m\neq T_H$ in relation with the Generalized Second Law, both in GR and $f(R)$). However we anticipate that when evaluating the entropy in an Hubble volume $V\sim H^{-3}$  and considering also the contribution of the horizon entropy \cite{eling2006non,Karami_2012,wu2012thermodynamic,cai2007unified},
\begin{equation}
T_H\frac{dS_{tot}}{dt}=T_H\frac{dS_{m}}{dt}+T_H\frac{dS_{H}}{dt},
\end{equation}
both in $f(R)$ and GR, the Generalized Second Law of thermodynamics (GSL) remains satisfied $\frac{dS_{tot}}{dt}\geq0$,where $T_H=\frac{H}{2\pi}\left(1+\frac{\dot{H}}{2H^2}\right)$ is the horizon temperature and 
\begin{equation}
    S_H=\frac{A f'(R)}{4G},
\end{equation}
is the horizon entropy \cite{Cognola:2012jz,PhysRevD.48.R3427}.  In the next subsections we will expand on \cite{Paul:2025rqe} allowing for irreversible processes (i.e. $\Delta S_h+\Delta S_m>0$) due to internal dissipative processes in the matter fluid even assuming $T_m=T_H$.\\
The \textit{total entropy} that enters the Generalized Second Law (GSL) combines matter with the geometric contribution of the apparent-horizon area.  Because horizon entropy grows, $d(S_{m}+S_{H})/dt\ge0$ can still hold even when $dS_{m}/dt<0$.  The GSL is therefore insensitive to the local inconsistency identified in the previous subsection and cannot be used to impose bounds on bulk viscosity in viscous $\Lambda$CDM model, and the net decrease of the matter entropy in an Hubble volume is not a symptom of ``ill behaved fluids'' as we are going to show explicitely.\\
 Taking, for example, an Hubble volume which is known, in the standard $\Lambda$CDM model, to be increasing, will lead to a net decrease in the entropy of dark matter $\frac{dS_m}{dt}<0$ even when it is assumed to be a perfect dust fluid. The reason lies in the interplay between the scaling of the  energy density $\rho_{dm}\sim a^{-3}$, and the volume $V\sim H^{-3}$, so the energy inside the horizon goes like $E\sim\dot{a}^{-3}$ and so the entropy $T\frac {dS}{dt}=\frac{dE}{dt}\sim-\frac{\ddot{a}}{\dot{a}^4}$, decreases since $\ddot{a}>0$, although it is known that perfect fluids do not violate the second law of thermodynamics locally. We conclude that, even though GSL arguments are valuable tools, they are not suitable in this case to highlight what are physically reasonable modification to the continuity equation. \\
Having clarified this point,  we now proceed to evaluate the entropy in a volume corresponding to the apparent horizon \cite{Karami_2012}, $r_H=(H^2+\frac{k}{a^2})^{-\frac{1}{2}}$, which corresponds to the Hubble radius in the case of spatially flat FLRW metric. We are going to assume the best fit values for the bulk viscosity obtained by \cite{Chi2025}, when the formal identification $2\eta\to-3\zeta$ \footnote{we remark that a non trivial bulk viscosity time dependence has been explained in \cite{Paul:2025rqe}.}  is assumed. Evaluating the variation of entropy for bulk viscous dust matter\footnote{Notice that for the cosmological constant $T_HdS_\Lambda=0$ so the only contribution to entropy changes comes from the matter sector and the horizon.} $T_HdS_m=d(\rho_{vdm} V)$, in a Hubble volume $V=\frac{4\pi}{3H^3}$, we obtain 
\begin{equation}
T_H\frac{dS_m}{dt}=\frac{4\pi}{3}\Big[\dot{\rho}_{vdm}\frac{1}{H^3}-3\rho_{vdm}\frac{\dot{H}}{H^4}\Big]=\frac{4\pi}{3}\Big[-3\rho_{vdm}\left(\frac{1}{H^2}+\frac{\dot{H}}{H^4}\right)+\frac{9\zeta}{H} 
 \Big],\label{matterentropy}
\end{equation}
where in the last equality we used  Eq.~\eqref{bulk}.
It's clear from Eq.~\eqref{matterentropy}, that to have an increasing entropy we have to require $3\zeta H\geq\rho_{vdm}\left(1+\frac{\dot{H}}{H^2}\right)$. Under the assumptions that $\zeta\geq0$, $H>0$  and $\rho_{vdm}>0$, the increasing entropy requirement becomes a requirement on the deceleration parameter $q$, $1+\frac{\dot{H}}{H^2}=-q$, hence $q\geq-\frac{3\zeta H}{\rho_{vdm}}$.  Indeed  the entropy in a Hubble volume may seem to increase for appropriate values of the bulk viscosity, however this condition is clearly violated for reasonable modifications of the $\Lambda$CDM model both in Einstein gravity and $f(R)$ as we are going to prove.  From the Friedmann equations with an effective pressure as the one assumed by \cite{Chi2025}, $p_{tot}=p_{eff}+p_{\Lambda}=-3\zeta H-\rho_{\Lambda 0}$, where we included a cosmological constant term, one finds that $\dot{H}$ depends on the shear viscosity\footnote{$ 2\dot{H}+3H^2=-\lambda_p^2(-\rho_{\Lambda 0}-3\zeta H)$}, the right hand side of  Eq.~\eqref{matterentropy} multiplied by $\frac{3H^2}{4\pi}$, can be rewritten as 
\color{black}
\begin{equation}
    3\rho_{vdm} q+9\zeta H=\frac{3 (\rho_{vdm} -2 \rho _{\Lambda 0}) (\rho_{vdm}-3 \zeta  H )}{2 (\rho_{vdm} +\rho_ {\Lambda 0})}\geq0,\label{cond}
\end{equation}
which in the reasonable case $\rho_{vdm}<2\rho_{\Lambda 0}$, is satisfied by $\zeta\geq\frac{\rho_{vdm}}{3H}>0$. We conclude that assuming the validity of Eq.~\eqref{41}, proviso the formal identification provided above, bulk viscosity should be positive in order to not violate the entropy increase for the matter sector, however in Table 1 of \cite{Chi2025}, they find a positive shear, which translates to a negative bulk, hence an entropy decrease for the matter sector.   
Moreover one can ask if the results given in the above mentioned Table 1, for the two models of viscosity they are assuming, are possibly compatible with an entropy increase when their lower $1\sigma$  deviation is assumed which  happens to be negative (hence a positive bulk). This is not the case since for the case $\eta=H\eta_v/\lambda_p^2$ which has to be correctly interpreted as $\zeta=H\zeta_v/\lambda_p^2$, the condition for entropy increase Eq.~\eqref{cond} simplifies to $\zeta_v\geq\Omega_{vdm}\approx0.32$, while they find $\eta_v\approx-5.7\times10^{-6}$, which translates to $\zeta_v\approx 3.8\times10^{-6}$, hence no matter entropy increase. For the case 2, they assumed a shear viscosity which goes like $\eta=\eta_v \frac{\rho_{vdm}}{H}$, which has to be interpreted as $\zeta=\zeta_v \frac{\rho_{vdm}}{H}$ which when inserted in the increasing entropy condition leads to $\zeta_v\geq\frac{1}{3}$, which is again violated even at the $1\sigma$ lower bound $\eta_v\approx-5.8\times 10^{-6}$, which implies $\zeta_v\approx3.87\times10^{-6}$.\\
We conclude the analysis of the entropy increase condition in GR with a final remark, the total entropy variation  is given by \cite{Karami_2012}
\begin{equation}
    T_H\frac{{dS}_{tot}}{dt}=\frac{H\dot{H}^2}{2G|H|^5}, \label{entropytotaleGR}
\end{equation}
hence always satisfies the GSL, for any type of matter contribution, as long as $H>0$. This makes, at least in the case of Einstein gravity, the GSL not suitable to constraint the viscosity of the dust matter fluid, as long as the geometry is described by a positive $H$. As a matter of fact, one can readily see that also a perfect dust fluid leads to a decrease in matter entropy, as one can see by setting $\zeta=0$ in Eq.~\eqref{cond}.  The requirement for increasing matter entropy becomes  $q>0$, which would lead to a decelerated expansion, in clear violation with the current observation of accelerated expansion \cite{Planck:2018vyg}.\\ 
\subsection{Generalized Second Law in $f(R)=R+\alpha R^2$}
Having acknowledged that a violation of the matter entropy increase in the case of GR is not necessarily a sign of ``ill-behaved" fluids, we evaluate directly the full entropy variation (horizon + matter) in the Starobinsky model, $f(R)=R+\alpha R^2$. From \cite{Karami_2012} we have for spatially flat FLRW geometry
\begin{equation}
T_H \frac{S_{tot}}{dt} = \frac{1}{4G |H|^5} \big[
2H \dot{H}^2 (1+2\alpha R)- 2\alpha\dot{H} H^2 \dot{R}
+ 4\alpha H (\dot{H} + H^2) \ddot{R}\label{startotentropy}
\big],
\end{equation}
which can be expanded in powers series of $\alpha$. Assuming $H=H^{(0)}+\alpha H^{(1)}$, one gets symbolically $T_H \frac{dS_{tot}}{dt}=T_H \frac{dS_{{tot}}^{GR}}{dt}+\alpha T_H \frac{ d\delta S_{tot}}{dt}$, where the first contribution is the one obtained in GR, and we already proved to be positive, while the second $\frac{d\delta S_{tot}}{dt}$  captures all the deviations introduced by the higher curvature correction. Since we expect the $\alpha$ correction to be perturbatively small compared to the leading order contribution  for perturbative reliability, we expect that also in this case the entropy is increasing as long as we are in the perturbative regime (as we are going to prove analytically in the next subsection). \\
For the matter entropy we expand in $\alpha$,  Eq.~\eqref{matterentropy} and we find
\begin{equation}
 \begin{aligned}
    T_H\frac{dS_m}{dt}&= \delta Q^{(0)}_m+\alpha\delta Q^{(1)}_m,\\
 \delta Q^{(0)}_m&\equiv\frac{4\pi}{3}\Big[-3\rho_{vdm}\left(\frac{1}{H_{(0)}^2}+\frac{\dot{H}^{(0)}}{H_{(0)}^4}\right)+9\frac{\zeta} {H^{(0)}} \Big],\\
 \delta Q^{(1)}_m&\equiv\frac{4\pi}{3}\Big[-3\rho_{vdm}\left(\frac{H^{(0)}\dot{H}^{(1)}-4H^{(1)}\dot{H}^{(0)}-2H_{(0)}^2H_{(1)}}{H_{(0)}^5}\right)-9\zeta \frac{H^{(1)}}{H_{(0)}^2}\Big],\label{matterenpert}
     \end{aligned}
 \end{equation}
 where we already proved that the first contribution  $\delta Q^{(0)}_m$ is negative, so for perturbative consistency one expects $\alpha|\delta Q^{(1)}_m|<|\delta Q^{(0)}_m|$, and so $T_H \frac{dS_m}{dt}<0$.  We will provide the explicit proof of the statement we made on general groundings with and without viscosity for the two case studies of \cite{Chi2025}.\\
\subsubsection{GSL in non viscous $f(R)$}
\begin{figure}[t!]
  \centering
  \includegraphics[width=8cm]{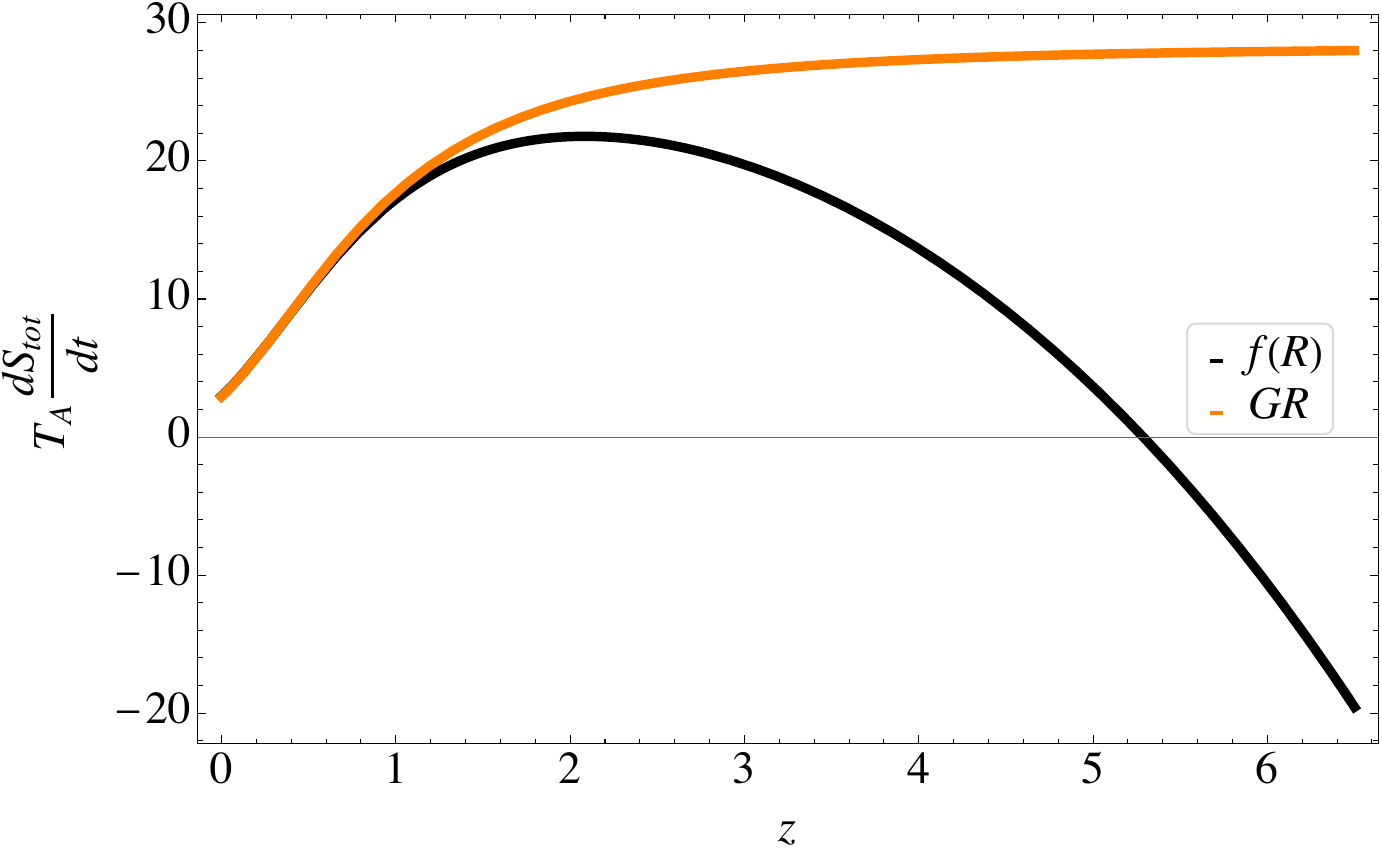}
  \caption{Time variation of the total entropy case 1, Matter+Horizon in $f(R)$ and GR ($\alpha=0$), $T_H\frac{dS_{\text{tot}}}{dt}$ as a function of the redshift $z$ for the mean values evaluated in \cite{Chi2025}, $\zeta_v=-1.79205\times10^{-6},\,H_0=75.42,\,\Omega_{vdm}=0.3216,\,\alpha=4.9\times10^{-8},\,\lambda_p=1,$ for $\zeta=H \zeta_v/\lambda_p^2$}
  \label{fig:entropy}
\end{figure}

\begin{figure}[t!]
  \centering
  \includegraphics[width=8cm]{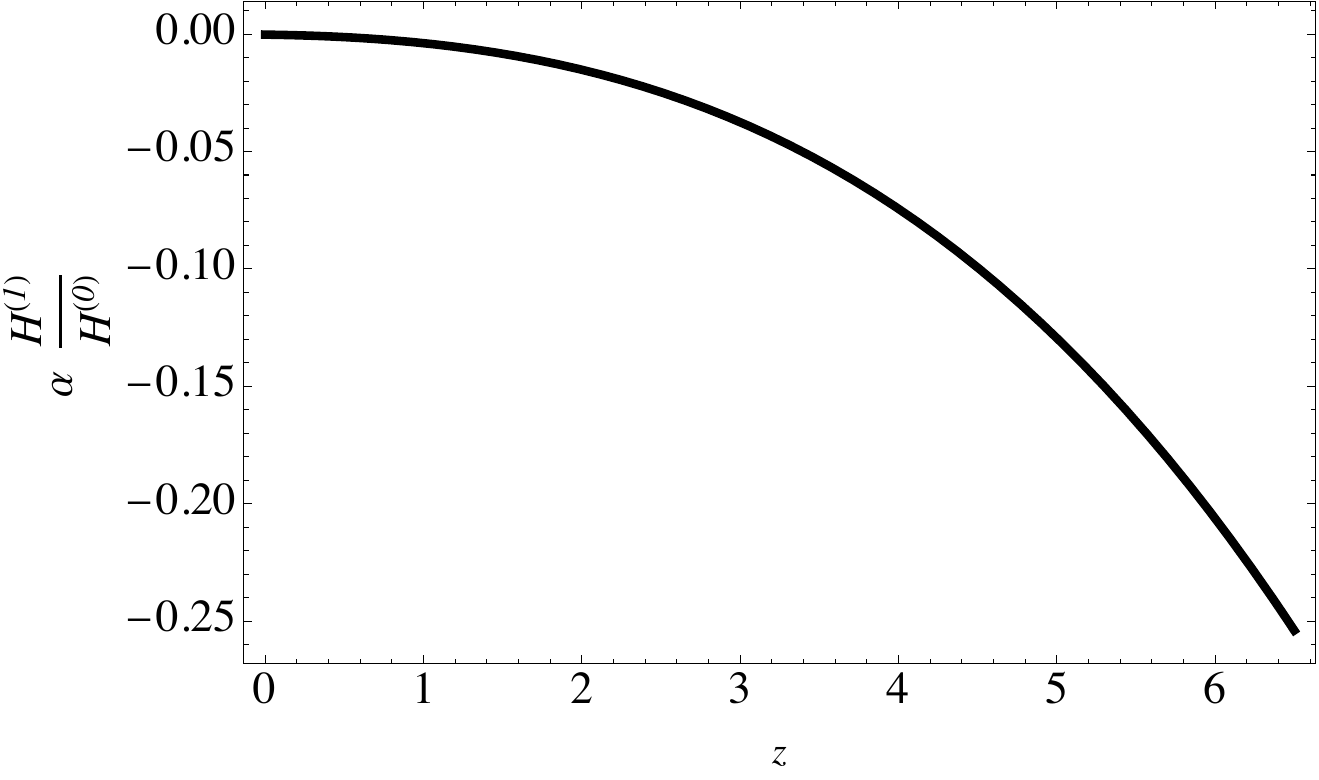}
  \caption{Ratio between the first order $\alpha H^{(1)}$,  and the zero order solution $H^{(0)}$  case 1 as a function of the redshift $z$ for the mean values evaluated in \cite{Chi2025}, $\zeta_v=-1.79205\times10^{-6},\,H_0=75.42,\,\Omega_{vdm}=0.3216,\,\alpha=4.9\times10^{-8},\,\lambda_p=1,$ for $\zeta=H \zeta_v/\lambda_p^2$}
  \label{fig:H1case1}
\end{figure}

 We start by analyzing the matter and total entropy evolution when there is no viscous contribution. We recall that the modified Friedmann equation reads \cite{Fanizza:2020hat} 
\begin{equation}
3H^2+18\alpha H^2 \Big[ 4(1+z)H H' - (1+z)^2\left( H'^2 + 2H H'' \right) \Big] = \lambda_P^2 \sum_i \rho_i(z), \label{general} 
\end{equation}
where $i$ labels the different fluid contributions (dark matter and cosmological constant), $z$ the redshift and with a $'$ the derivative with respect to $z$. Upon the insertion of the perturbative ansatz $H=H^{(0)}+\alpha H^{(1)}$, it can be solved  together with the continuity equation Eq.~\eqref{continuitycorrect} to give
\begin{align}
    H^{(0)}&=H_0\left(\Omega_{dm0}(1+z)^3+\Omega_{\Lambda0}\right)^{\frac{1}{2}},\\ \nonumber
    H^{(1)}&=-\frac{27}{4}H_0^3\Omega_{dm0}^2(1+z)^6\left(\Omega_{dm0}(1+z)^3+\Omega_{\Lambda0}\right)^{-\frac{1}{2}}.
\end{align}
Inserting these relations into Eq.~\eqref{matterenpert} we get the matter entropy variation  without shear viscosity
\begin{equation}
\begin{aligned}
T_H\frac{dS_m}{dt}&=\frac{4\pi}{3}\Big[\frac{9\, \Omega_{dm0}\, (1+z)^3 \left[ \Omega_{dm0}\, (1+z)^3 - 2 \Omega_{\Lambda0} \right]}{2 \lambda_p^2 \left( \Omega_{\Lambda0} + \Omega_{dm0}\, (1+z)^3 \right)^2} \\
&- \frac{243\, \alpha\, H_0^2\, \Omega_{dm0}^3\, (1+z)^9 \left( 4 \Omega_{\Lambda0} + \Omega_{dm0}\, (1+z)^3 \right)}{2 \lambda_p^2 \left( \Omega_{\Lambda0} + \Omega_{dm0}\, (1+z)^3 \right)^3}\Big].
\end{aligned}
\end{equation}
The  equation above shows that regardless of the presence of bulk viscosity, for sufficiently high values of the redshift, the matter entropy variation becomes negative, since the first term (which is not affected by $\alpha$ corrections) approaches a constant value while the $\alpha$ term decreases like $z^3$, more precisely for $z\gg1$
\begin{equation}
   T_H\frac{dS_m}{dt}\sim \frac{6\pi}{\lambda_P^2} \left[
1 - 27\, \alpha\, H_0^2\, \left( 
\Omega_{dm0} (1+z)^3 + \Omega_{\Lambda0}
\right)
\right].\label{asmat}
\end{equation}
The same applies to the total entropy variation Eq.~\eqref{startotentropy} ($G=\frac{\lambda_p^2}{8\pi}$)
\begin{equation}
\begin{aligned}
T_H\frac{dS_{tot}}{dt}&=\frac{9\pi\, \Omega_{dm0}^2\, (1+z)^6}{\lambda_p^2 \left( \Omega_{\Lambda0} + \Omega_{dm0}\, (1+z)^3 \right)^2} - \frac{27\pi\, \alpha\, H_0^2\, \Omega_{dm0}\, (1+z)^3}{\lambda_p^2 \left( \Omega_{\Lambda0} + \Omega_{dm0}\, (1+z)^3 \right)^3}\\
&\times \bigg[
-8\, \Omega_{\Lambda0}^3 + 15\, \Omega_{dm0}^3\, (1+z)^9+ 10\, \Omega_{dm0}^2\, \Omega_{\Lambda0}\, (1+z)^6  - 22\, \Omega_{dm0}\, \Omega_{\Lambda0}^2\, (1+z)^3
\bigg],
\end{aligned}
\end{equation}
which again has a similar asymptotic behavior
\begin{equation}
T_H\frac{dS_{tot}}{dt}\sim \frac{9\pi}{\lambda_p^2} \left[ 1-15\, \alpha\, H_0^2 \left( 3\, \Omega_{dm0}\, (1+z)^3 - 7\, \Omega_{\Lambda0} \right)  \right].\label{astot}
\end{equation}
From Eq.~\eqref{asmat}  and Eq.~\eqref{astot}, we can estimate at which redshift $z_d$, the $\alpha$ correction in the entropy evolution becomes of the same order of the leading order term, which can give a rough estimate of at which redshift the entropy starts decreasing. In both cases we find $1+z_d\sim\mathcal{O}(10^{-1})(\alpha H_0^2\Omega_{dm0})^{-\frac{1}{3}}$, where the $\mathcal{O}(10^{-1})$ number is approximately $0.3$. It is immediate to see that at this redshift at the  zeroth order in $\alpha$,  the ratio $\alpha\lvert\frac{H^{(1)}}{H^{(0)}}\rvert\sim\frac{27}{4}\alpha H_0^2\Omega_{dm0}(1+z_d)^3\sim\mathcal{O}(10^{-1})$, where the $\mathcal{O}(10^{-1})$ coefficient is approximately $0.18$, hence the reliability of the perturbative expansion fails. This argument does not involve viscosity hence it is general, \textit{Entropy starts to decrease for the breakdown of the perturbative approach}. We can conclude that the high $z$ negative value for the time variation of the total entropy and matter entropy Figs.~\ref{fig:entropy}, \ref{fig:matterentropycase1}, \ref{fig:entropy2} and \ref{fig:matterentropycase2} are not due to the presence of the bulk viscosity contribution but  due to the breakdown of the perturbative expansion reliability as can be seen in Figs.\ref{fig:H1case1} and \ref{fig:H1case2} where at high $z$ the correction $\alpha \lvert\frac{H^{(1)}}{H^{(0)}}\rvert\sim\mathcal{O}(10^{-1})$. \\
In the following subsections we give a quantitative analysis of the statement that the high $z$ entropy decrease is due to the breakdown of the perturbative approach and it is not due to viscous contribution and that the matter entropy increase condition on viscosity for $z=0$ is worsen by the presence of $\alpha$ corrections, and this is a general statement that is true for both the science cases of \cite{Chi2025}. Moreover we show the asymptotic behaviors for $z=0$ and $z\gg 1$ of the matter and total entropy variation in the two science cases, ensuring that with and without viscosity, the matter entropy is decreasing today ($z=0$) regardless of the presence of  viscosity, proving also in the context of Starobinsky gravity that GSL is not suitable to constraint the bulk viscosity from a thermodynamical standpoint. Since the matter entropy is decreasing even without bulk viscosity, this reinforce our argument made for the GR case, where the matter entropy decrease is present also for a simple dust fluid. Finally in the case $z\gg 1$, we prove that the correction introduced by the viscosity does not lead to entropy decrease and that the decreasing entropy is solely related to the $\alpha$ terms in Eqs.~\eqref{asmat} and \eqref{astot} and not on the bulk viscosity at the leading order in $\alpha$ and $\zeta$. 
\subsection{Case 1 $\zeta=H \zeta_v/\lambda_p^2$}
\begin{figure}[t!]
  \centering
  \includegraphics[width=8cm]{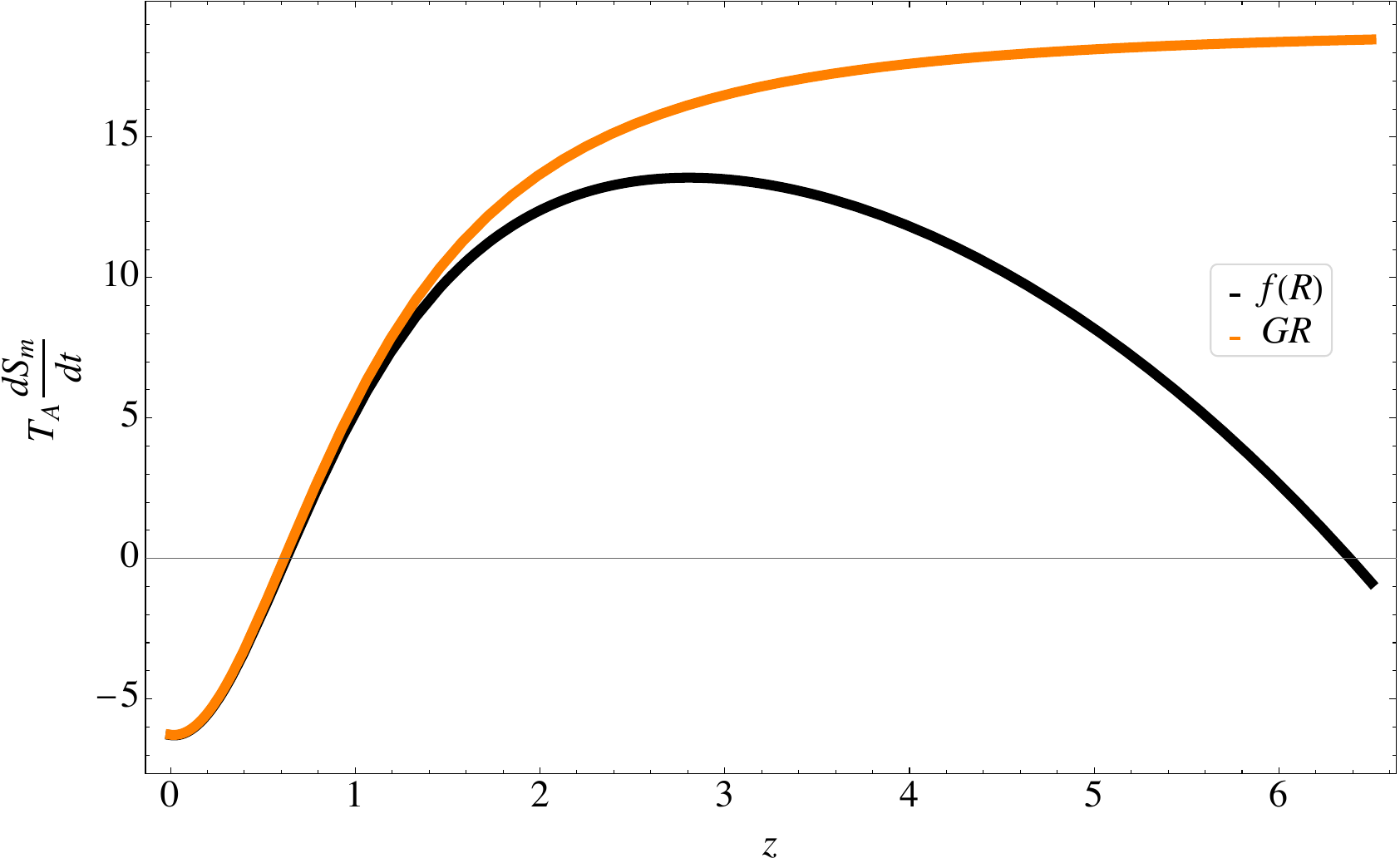}
  \caption{Time variation of the matter entropy case 1, $T_H\frac{dS_m}{dt}$ in  $f(R)$ and GR ($\alpha=0$) as a function of the redshift $z$ for the mean values evaluated in \cite{Chi2025}, $\zeta_v=-1.79205\times10^{-6},\,H_0=75.42,\,\Omega_{vdm}=0.3216,\,\alpha=4.9\times10^{-8},\,\lambda_p=1$, for $\zeta=H \zeta_v/\lambda_p^2$}
  \label{fig:matterentropycase1}
\end{figure}

Solving perturbatively Eq.~\eqref{general}  together with Eq.~\eqref{bulk} in this science case we have
\begin{equation}
    \rho_{vdm}=\rho_{vdm0}(1+z)^{3(1-\zeta_v)}-\zeta_v\frac{(1+z)^{3(1-\zeta_v)}-1}{1-\zeta_v}\rho_{\Lambda0}, \label{rhocase1} 
\end{equation}
and inserting this result into Eq.~\eqref{general} we get
\begin{equation}
H^{(0)}=H_0\Big[\Omega_{vdm0}(1+z)^{3(1-\zeta_v)}+\left(1-\zeta_v\frac{(1+z)^{3(1-\zeta_v)}-1}{1-\zeta_v}\right)\Omega_{\Lambda0}\Big]^{\frac{1}{2}}.\label{case1H0}
\end{equation}
 From Eq.~\eqref{general} we also find the correction to the Hubble function as
\begin{equation}
\begin{aligned}
H^{(1)}&= -\frac{
9 H_0^3\, (1+z)^{3(1 - \zeta_v)}\, 
\left[ (\zeta_v - 1)\, \Omega_{vdm0} + \zeta_v\, \Omega_{\Lambda0} \right]
}{
4(1 - \zeta_v)
} \\
&  \times
\frac{
 3(3\zeta_v + 1)\, (1+z)^{3(1 - \zeta_v)}\, 
\left[ (\zeta_v - 1)\, \Omega_{vdm0} + \zeta_v\, \Omega_{\Lambda0} \right] 
- 12\zeta_v\, \Omega_{\Lambda0} 
}{
\sqrt{
\frac{
\Omega_{\Lambda0} - (1+z)^{3(1 - \zeta_v)}\, 
\left[ (\zeta_v - 1)\, \Omega_{vdm0} + \zeta_v\, \Omega_{\Lambda0} \right]
}{
1 - \zeta_v
}
}
}.\label{case1H1}
\end{aligned}
\end{equation}
We note that eqs.(42),(50) and (51), (i.e. $H^{(0)}(z)$ and $H^{(1)}(z)$) as a function of the bulk viscous parameter $\zeta_v$ provided in $\cite{Chi2025}$, are not consistent with the Friedmann equations (we remind to the reader that these results have been checked upon the formal substitution $2\eta_v\to-3\zeta_v$). This is a critical point that poses some doubt with the consistency of the parameter estimation of \cite{Chi2025}.\\
Having clarified this point, we proceed to evaluate $T_H\frac{dS_m}{dt}$ for $z=0$, inserting Eqs.~\eqref{case1H0} and \eqref{case1H1} into Eq.\eqref{matterenpert}, and neglecting terms $\mathcal{O}(\alpha \,H_0^2 \,\zeta_v)$ which are higher order in the perturbative expansion we find
\begin{equation}
\begin{aligned}
T_H\frac{dS_m}{dt}&\sim\frac{6\pi\, (\Omega_{vdm0} - 2\, \Omega_{\Lambda0})\, \left (  \Omega_{vdm0}-\zeta_v\right) }{\lambda_P^2}- \frac{162\, \pi\, \alpha\, H_0^2\, \Omega_{vdm0}^3\, (\Omega_{vdm0} + 4\, \Omega_{\Lambda0})}{\lambda_P^2\,}
\end{aligned}
\end{equation}
where the first term exactly matches the one we found in our Eq.~\eqref{cond} up to a factor $\frac{4\pi}{3}$  and a multiplication by $H^2$, hence recovering the GR case in the limit $\alpha\to0$ as expected. From the equation above, it is clear that the presence of the $\alpha$ correction does not ameliorate the condition for entropy increase, on the contrary it worsen the bound for the bulk viscosity that should be even more positive. As a matter of fact without $\alpha$, we should demand $\zeta_v\geq\Omega_{vdm0}$, as already proven in the GR case, but in Starobinsky
\begin{equation}
   \zeta_v\geq\Omega_{vdm0}
+ 27\,\alpha\, H_0^2\frac{\, \, \Omega_{vdm0}^3\, (\Omega_{vdm0} + 4\, \Omega_{\Lambda0})}{ (2\, \Omega_{\Lambda0}-\Omega_{vdm0} )},
\end{equation}
where the $\alpha$ term is positive, hence $\zeta_v$ should be even more positive with respect to the GR case to ensure matter entropy increase. We notice however that the matter entropy decrease is present also for perfect dust fluid, where $\zeta_v=0$, hinting that this is not in general a sign of ill behaved fluids.  \\
For what concerns the behavior of the total entropy variation at $z=0$ with bulk viscosity, (neglecting terms $\mathcal{O}(\alpha\,H_0^2\,\zeta_v)$), we find
\begin{equation}
    T_H\frac{dS_{tot}}{dt}\sim\frac{9\pi\, \left(  \Omega_{vdm0}-\zeta_v  \right)^2}{\lambda_P^2}- \frac{27\, \pi\, \alpha\, H_0^2\, \Omega_{vdm0}}{\lambda_p^2}\times \left(
15\, \Omega_{vdm0}^3 + 10\, \Omega_{vdm0}^2\, \Omega_{\Lambda0} - 22\, \Omega_{vdm0}\, \Omega_{\Lambda0}^2 - 8\, \Omega_{\Lambda0}^3
\right).\label{totcase1}
\end{equation}
From the previous equation we find that the first term is always positive in concordance with the general results of the GR case of the previous section, and that the additional terms in $\alpha$ may lead to a potential violation of the GSL. Noticing that the $\alpha$ correction is  positive for $0\leq\Omega_{vdm0}\lesssim0.52$, we conclude that the previous expression is always positive for reasanoble values of the cosmological parameters.
We deduce that in Starobinsky gravity whether there is a perturbative viscous contribution or not, a  dust fluid violates the entropy increase of matter, but the total entropy always increases (at least for $z=0$) underlying once more that GSL is not a valuable tool in this context.\\
We finally conclude evaluating what is the contribution to the matter entropy  and total entropy variation at $z\gg1$, when viscosity is taken into account. What we find corroborates the results already stated in the previous section, the breakdown of the perturbative approach is responsible for the  matter and total entropy decrease at large redhsifts. At the first order in $\zeta_v$ and neglecting $\mathcal{O}(\alpha\,H_0^2\,\zeta_v)$, for $z\gg1$ the entropy variations are given by
\begin{align}
T_H\frac{dS_{tot}}{dt}&\sim \frac{9\pi}{\lambda_p^2} \left[ 1-2\zeta_v-15\, \alpha\, H_0^2 \left( 3\, \Omega_{vdm0}\, (1+z)^3 - 7\, \Omega_{\Lambda0} \right)  \right],\label{case1totinftot} \\ 
   T_H\frac{dS_m}{dt}&\sim \frac{6\pi}{ \lambda_P^2} \left[
1 -\zeta_v- 27\, \alpha\, H_0^2\, \left( 
\Omega_{vdm0} (1+z)^3 + \Omega_{\Lambda0}
\right)
\right] \label{case1totinfmat}.
\end{align}
  What we find is that all the considerations of the case without  viscosity still apply. As a matter of fact the decreasing terms are controlled by $\alpha$, as before, imposing in Eq.~\eqref{case1totinftot} or in Eq.~\eqref{case1totinfmat} that the decreasing term is of the same order of the leading term, we find up to terms in $\mathcal{O}(\alpha\,H_0^2\,\zeta_v)$, $1+z_d\sim\mathcal{O}(10^{-1})(\alpha H_0^2\Omega_{vdm0})^{-\frac{1}{3}}$, so the same result as the non viscous case follows, $\alpha\lvert\frac{H^{(1)}}{H^{(0)}}\rvert\sim\frac{27}{4}\alpha H_0^2\Omega_{dm0}(1+z_d)^3+\mathcal{O}(\alpha\,H_0^2\,\zeta_v)z^3\log(z)\sim\mathcal{O}(10^{-1})$. As can be seen in Fig.\ref{fig:matterentropycase1} , also in this case the matter entropy decreases at low redshift, and  as in the GR ($\alpha=0$) case, this behavior is manifest also for simple non viscous dust matter ($\zeta_v=0$), making the GSL not an instructive tool to constraint possible bulk viscous contributions. We point out that for Figs. \ref{fig:entropy}, \ref{fig:H1case1} and \ref{fig:matterentropycase1}, we used the fiducial values given in \cite{Chi2025} with the formal identification $2\eta_v\to-3\zeta_v$.\\
  Indeed the transition from the decreasing behavior to the increasing behavior in Figs.\ref{fig:matterentropycase1} and \ref{fig:matterentropycase2}, happens at the transition between CDM and $\Lambda$ domination $z^*\approx 0.7$, and so when the cosmological evolution becomes accelerated and so $q$ becomes negative, leading to a decrease in the matter entropy as in the case of GR in concordance with the discussion made after Eq.~\eqref{matterentropy} with non dissipative dust matter.
\subsection{Case 2 $\zeta=\rho_{vdm} \zeta_v/H$}
\begin{figure}[t!]
  \centering
  \includegraphics[width=8cm]{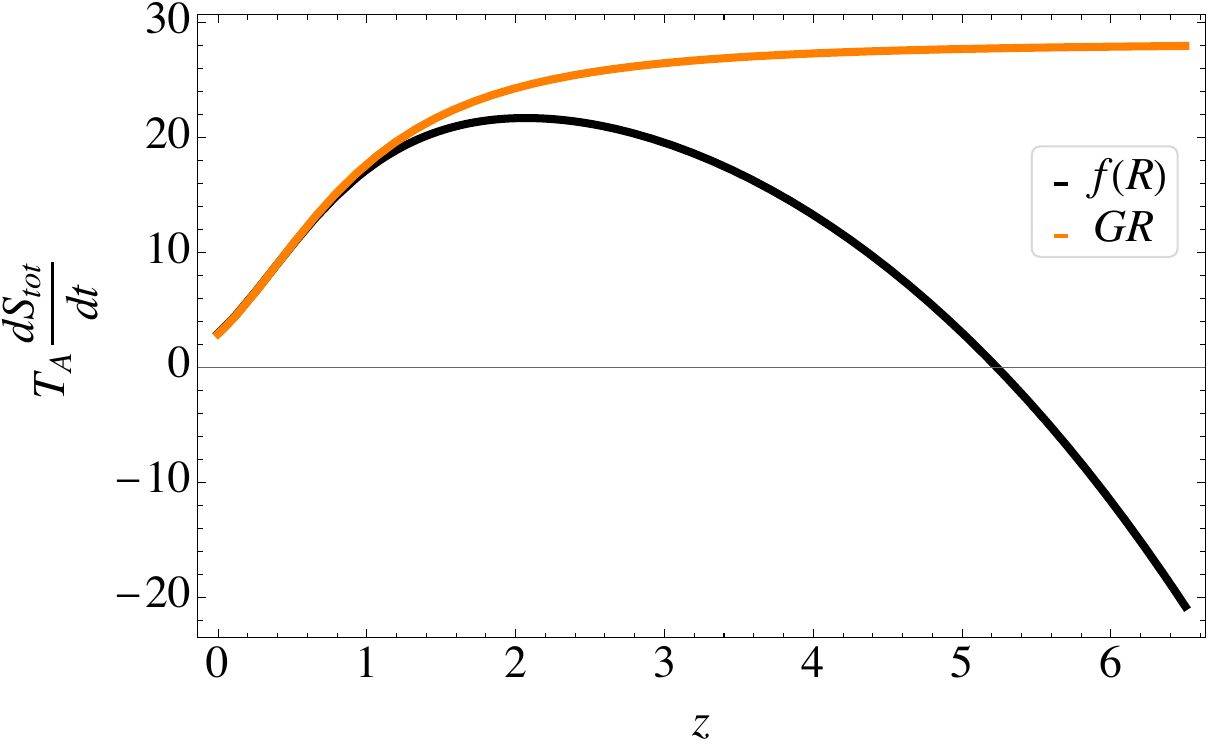}
  \caption{Time variation of the total entropy case 2, Matter+Horizon in $f(R)$ and GR ($\alpha=0$), $T_H\frac{dS_{\text{tot}}}{dt}$ as a function of the redshift $z$ for the mean values evaluated in \cite{Chi2025}, $\zeta_v=-2.5038\times10^{-6},\,H_0=75.40,\,\Omega_{vdm}=0.3222,\,\alpha=5.02\times10^{-8},\,\lambda_p=1,$ for $\zeta= \zeta_v\frac{\rho_{vdm}}{H}$}
  \label{fig:entropy2}
\end{figure}

\begin{figure}[t!]
  \centering
  \includegraphics[width=8cm]{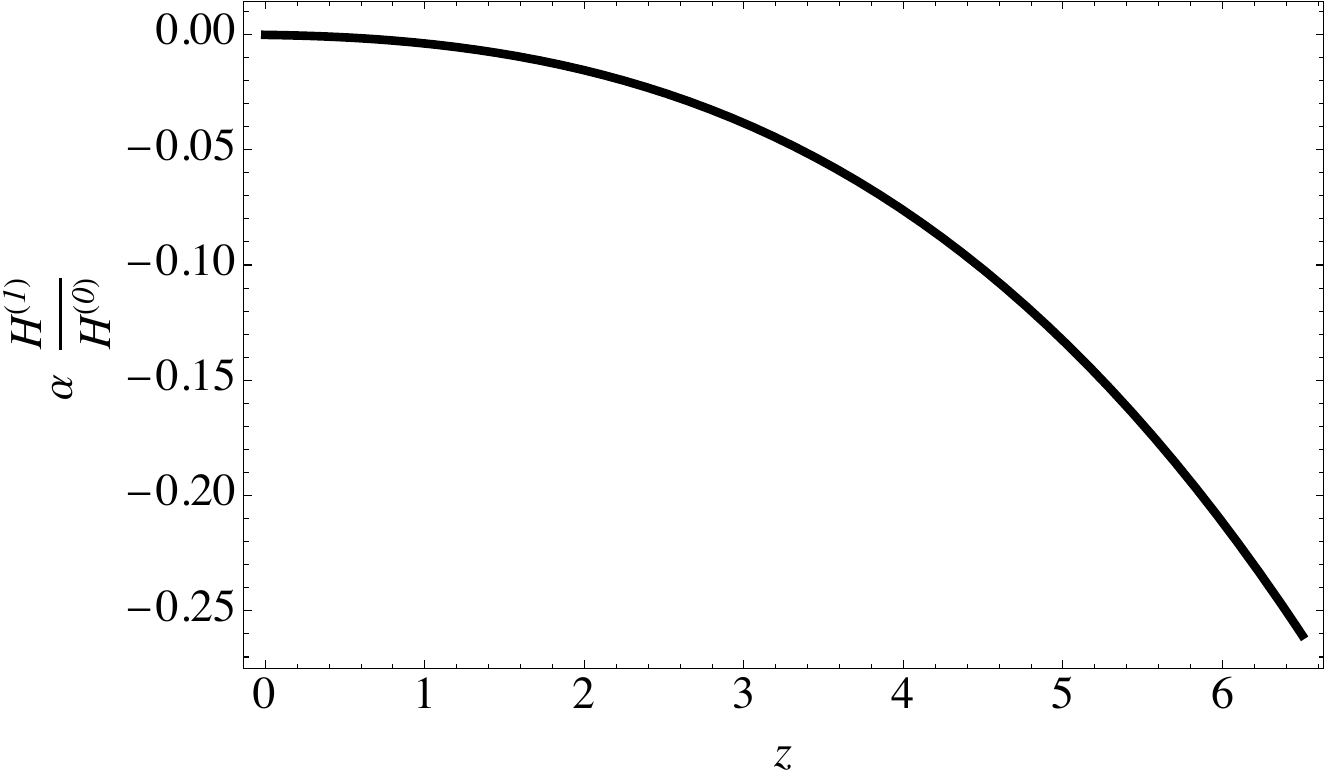}
  \caption{Ratio between the first order $\alpha H^{(1)}$ and the zero order solution $H^{(0)}$, case 2  as a function of the redshift $z$ for the mean values evaluated in \cite{Chi2025}, $\zeta_v=-2.5038\times10^{-6},\,H_0=75.40,\,\Omega_{vdm}=0.3222,\,\alpha=5.02\times10^{-8},\,\lambda_p=1,$ for $\zeta= \zeta_v\frac{\rho_{vdm}}{H}$}
  \label{fig:H1case2}
\end{figure}

\begin{figure}[t!]
  \centering
  \includegraphics[width=8cm]{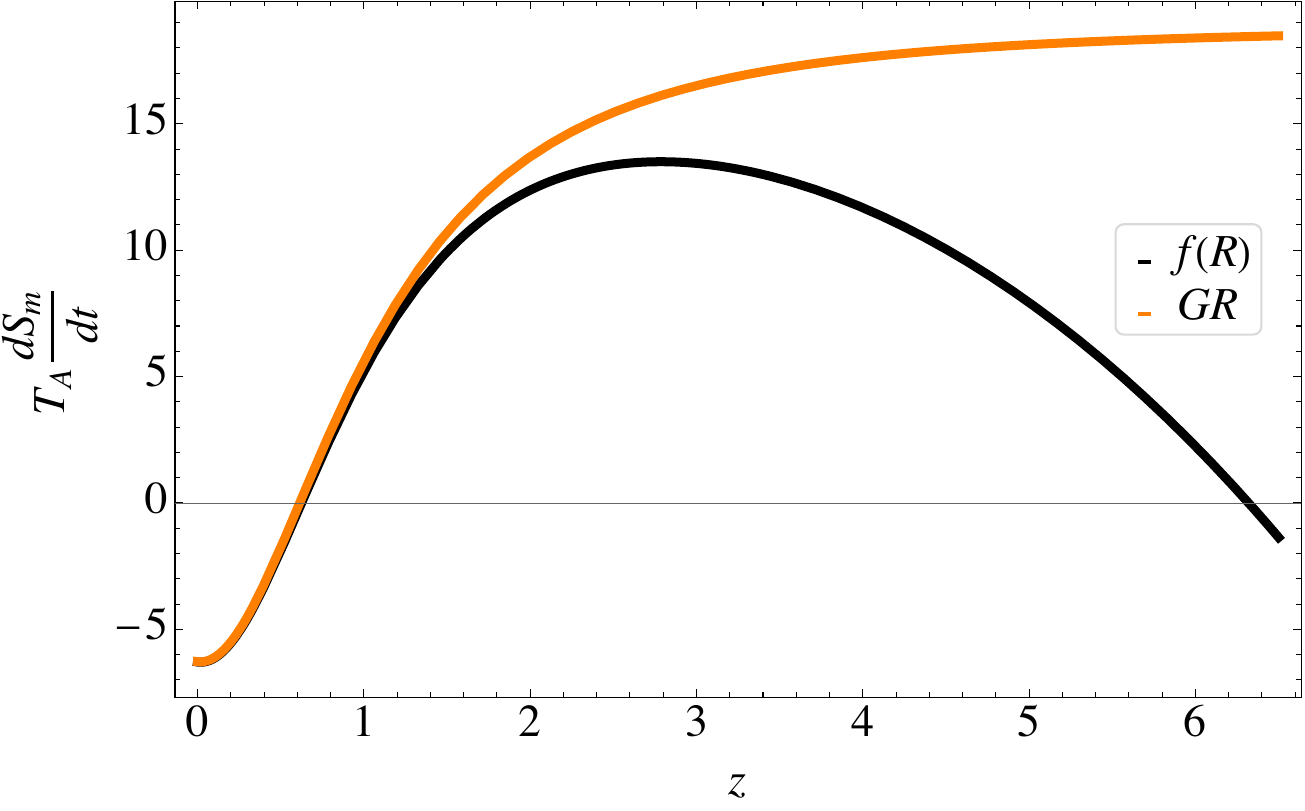}
  \caption{Time variation of the matter entropy case 2, $T_H\frac{dS_{m}}{dt}$ in  $f(R)$ and GR ($\alpha=0$) as a function of the redshift $z$ for the mean values evaluated in \cite{Chi2025}, $\zeta_v=-2.5038\times10^{-6},\,H_0=75.40,\,\Omega_{vdm}=0.3222,\,\alpha=5.02\times10^{-8},\,\lambda_p=1,$ for $\zeta= \zeta_v\frac{\rho_{vdm}}{H}$}
  \label{fig:matterentropycase2}
\end{figure}

In this science case the solution to the Friedmann equation Eq.~\eqref{general} and the continuity equation Eq.~\eqref{bulk} at the lowest order reads 
\begin{equation}
H^{(0)}(z) = H_0 \left[ \Omega_{vdm0}(1+z)^{3(1-3\eta_v)} + \Omega_{\Lambda0} \right]^{1/2}, \label{case2H0}
\end{equation}
and the first order $\alpha$ correction is
\begin{equation}
\begin{aligned}
H^{(1)}(z) =\;& -\frac{27}{4} H_0^3\, (1+z)^{3(1 - 3\zeta_v)}\, (1 - 3\zeta_v)\, \Omega_{vdm0} \notag \times \left[ (1 + 9\zeta_v)\, (1+z)^{3(1 - 3\zeta_v)}\, \Omega_{vdm0} + 12\zeta_v\, \Omega_{\Lambda0} \right] \notag \\
&\times \left[ \Omega_{vdm0}\, (1+z)^{3(1 - 3\zeta_v)} + \Omega_{\Lambda0} \right]^{-1/2} \label{case2H1}
\end{aligned}
\end{equation}
Evaluating $T_H\frac{dS_m}{dt}$ for $z=0$  and neglecting terms $\mathcal{O}(\alpha\,H_0^2 \zeta_v)$ which are higher order in the perturbative expansion we have
\begin{equation}
T_H\frac{dS_m}{dt}\sim\frac{6\pi\, \Omega_{vdm0}( 1-3\zeta_v)\,  \left( \Omega_{vdm0} - 2\, \Omega_{\Lambda0} \right)}{\lambda_P^2\,}- \frac{162\, \pi\, \alpha\, H_0^2\, \Omega_{vdm0}^3 \left( \Omega_{vdm0} + 4\, \Omega_{\Lambda0} \right)}{\lambda_P^2\,},
\end{equation}
where again the first term exactly reproduces the one in Eq.~\eqref{cond} up to a factor $\frac{4\pi}{3}H^2$, so we recover the GR case in the limit $\alpha\to0$ as expected. Also in the case study 2 we find that the presence of the $\alpha$ correction does not ameliorate the condition for entropy increase. We find that without $\alpha$, we should demand $\zeta_v\geq\frac{1}{3}$, as already proven in the GR case in the previous section, but in $f(R)$
\begin{equation}
\zeta_v \geq  \frac{1}{3}+9\alpha\,H_0^2\frac{ \Omega_{vdm0}^2\, (\Omega_{vdm0} + 4\, \Omega_{\Lambda0})}
{\, ( 2\, \Omega_{\Lambda0}-\Omega_{vdm0} )\, (\Omega_{vdm0} + \Omega_{\Lambda0})}.
\end{equation}
For what concerns the behavior of the total entropy variation at $z=0$ (neglecting terms $\mathcal{O}(\alpha\,H_0^2\,\zeta_v)$), we find
\begin{equation}
\begin{aligned}
T_H\frac{dS_{tot}}{dt}&\sim\frac{9\pi\,\Omega_{vdm0}^2 \left(1-3\zeta_v \right)^2}{\lambda_P^2}- \frac{27\pi\, \alpha\, H_0^2\, \Omega_{vdm0}}{\lambda_p^2}\\
& \times\left(
15\, \Omega_{vdm0}^3 + 10\, \Omega_{vdm0}^2\, \Omega_{\Lambda0} - 22\, \Omega_{vdm0}\, \Omega_{\Lambda0}^2 - 8\, \Omega_{\Lambda0}^3
\right).
\end{aligned}
\end{equation}
We notice that also in this case, the total entropy is always increasing at the zeroth order in  $\alpha$ and since the $\alpha$ correction are the same as the one in Eq.~\eqref{totcase1}, the same argument applies as in the previous case, namely the total entropy is always increasing for $0\leq\Omega_{vdm0}\lesssim0.52$.
Also in this case for vanishing viscosity, we find that a  dust fluid violates the entropy increase of matter, but the total entropy always increases (at least for $z=0$).\\
We conclude evaluating what is the contribution to the matter entropy  and total entropy variation at $z\gg1$, in this second case, although the conclusion we draw are the same as the one of case 1.  We find that at the first order in $\zeta_v$ and neglecting $\mathcal{O}(\alpha\,H_0^2\,\zeta_v)$, for $z\gg1$
\begin{align}
T_H\frac{dS_{tot}}{dt}&\sim \frac{9\pi}{\lambda_p^2} \left[ 1-6\zeta_v-15\, \alpha\, H_0^2 \left( 3\, \Omega_{vdm0}\, (1+z)^3 - 7\, \Omega_{\Lambda0} \right)  \right],\label{case2totinftot} \\ 
   T_H\frac{dS_m}{dt}&\sim \frac{6\pi}{\lambda_P^2} \left[
1 -3\zeta_v- 27\, \alpha\, H_0^2\, \left( 
\Omega_{vdm0} (1+z)^3 + \Omega_{\Lambda0}
\right)
\right] \label{case2totinfmat}.
\end{align}
 Up to some numerical coefficients these expressions match the one obtained in case 1, hence the same results apply $\alpha\lvert\frac{H^{(1)}}{H^{(0)}}\rvert\sim\frac{27}{4}\alpha H_0^2\Omega_{vdm0}(1+z_d)^3+\mathcal{O}(\alpha\,H_0^2\,\zeta_v)z^3\log(z)\sim\mathcal{O}(10^{-1})$.\\
 We conclude that the decreasing entropy is not due to bulk viscosity, but due to the breakdown of the perturbative reliability in the $\alpha$ expansion.  We show in Figs.~\ref{fig:entropy2}, \ref{fig:H1case2} and \ref{fig:matterentropycase2}, the plots for the time variation of the total entropy, the matter entropy and the ratio $\alpha\frac{H^{(1)}}{H^{(0)}}$ with the cosmological parameters estimated in \cite{Chi2025}, that corroborates the analytical results obtained. \\
\newpage
\color{black}
\section{Conclusions}

In this work, we have shown that shear viscosity does not influence the background dynamics of isotropic cosmological models, as long as the fluid is comoving and geodesic. Our analysis relies on general geometric arguments valid for a broad class of isotropic (but not necessarily homogeneous) metrics, and the result holds independently of the gravitational theory under consideration. In particular, we verified that in both General Relativity and $f(R)$ gravity including Starobinsky model the shear tensor vanishes identically for comoving fluids in these backgrounds, implying that anisotropic stresses do not enter the continuity equation.

We extended this result to both Jordan and Einstein frames, showing that conformal transformations do not introduce shear-dependent terms into the energy conservation equation when starting from a shear-free configuration. This supports the conclusion that shear viscosity cannot alter the evolution of the Hubble parameter, and thus cannot affect the electromagnetic luminosity distance in FLRW, which depends only on the background geometry via the Hubble parameter.

These findings have been compared with the assumptions made in \cite{Chi2025}, where shear viscosity was treated as a source modifying the continuity equation and thus the expansion history. We have shown that the continuity equation used there can only be recovered formally by reinterpreting the shear contribution as an effective bulk viscosity term. However, the positive values obtained for the shear viscosity in their fits would correspond to negative bulk viscosity, which violates the second law of thermodynamics in comoving volumes.

We also examined the entropy evolution in both comoving and Hubble volumes, showing that a decrease in matter entropy in an Hubble volume can arise even in the absence of viscosity, particularly in accelerated expansion phases. Although the Generalized Second Law (GSL) remains valid as the horizon entropy increases it does not provide useful constraints on the physical consistency of viscous modifications. Furthermore, we showed that the apparent entropy decrease at high redshift in viscous Starobinsky model is driven by the breakdown of the perturbative expansion in $\alpha$, rather than by dissipative effects.

Overall, while our results do not exclude the possibility that shear viscosity plays a role in more complex or perturbed settings (e.g. tensor modes), they suggest that constraints on shear viscosity cannot be reliably extracted from background observables such as the electromagnetic luminosity distance alone. Instead, multi-messenger observations which allow for a direct comparison between gravitational and electromagnetic distances remain the most robust avenue to probe dissipative properties of the dark sector, as previously proposed in \cite{Fanizza_2022}.
\\
Future studies will address the influence of viscosity (both bulk and shear) in the string based pre-Big Bang scenario \cite{{Gasperini:1992em},{Gasperini:2002bn},{Gasperini:2007}}, where a dissipative gas of string-holes is expected to naturally arise in the high curvature regime \cite{{Veneziano:2003sz},{PhysRevD.105.023532},{Bitnaya:2023vda}}, on the spectra of primordial gravitational waves (see \cite{{Conzinu:2024cwl},{Ben-Dayan:2024aec},{Conzinu:2025sot}} for recent discussions), on possible Primordial Black Hole production \cite{{Conzinu:2020cke},Conzinu:2023fui} and on the bouncing transition \cite{{Gasperini:2023tus},{Conzinu:2023fth}} from the contracting pre Big-Bang phase to the standard radiation epoch of the cosmological concordance model.

\section*{Acknowledgements}
EP and LT are supported in part by INFN under the program TAsP: \textit{``Theoretical Astroparticle Physic''}. EP and LT are also supported by the research grant number 2022E2J4RK ``PANTHEON: Perspectives in Astroparticle and Neutrino THEory with Old and New messenger'', under the program PRIN 2022 funded by the Italian Ministero dell'Universit\`a e della Ricerca (MUR) and by the European UnionNext Generation EU. The authors  want  to  acknowledge  the  hospitality  of  CERN,  Department  of Theoretical Physics.

\bibliography{biblio}
\bibliographystyle{JHEP}

\end{document}